%% file: main.tex
\documentclass[floatfix,aps,prl,reprint,superscriptaddress]{revtex4-2}
\usepackage{palatino}
\usepackage{graphicx}
\usepackage{amssymb}
\usepackage{amsmath}

\usepackage{xr-hyper} 
\usepackage[hidelinks]{hyperref} 
\usepackage{cleveref} 

\Crefname{figure}{Figure}{Figures}
\crefname{figure}{Fig.}{Figs.}
\Crefname{equation}{Equation}{Equations}
\crefname{equation}{Eq.}{Eqs.}
\Crefname{section}{Section}{Sections}
\crefname{section}{Sec.}{Secs.}

\renewcommand\rm[1]{{\mathrm{#1}}}

\usepackage{bm} 
\usepackage{mathrsfs} 
\usepackage{mathtools}
\usepackage{tikz}
\usepackage{todonotes}
\usepackage{url}
\usepackage{lipsum}
\usepackage{pgffor}
\usepackage{physics}
\usepackage{float}

\input{circuits.tex}

\DeclareFontEncoding{LS1}{}{}
\DeclareFontSubstitution{LS1}{stix}{m}{n}
\DeclareSymbolFont{stixletters}{LS1}{stix}{m}{it}
\DeclareMathAccent{\vecev}{\mathord}{stixletters}{"95}

\usetikzlibrary{calc}
\usetikzlibrary{matrix}

\newcommand\dts{{..}} 
\newcommand\cdts{{\mkern 0mu\cdot \mkern 0mu \cdot}} 
\newcommand\sbra[1]{\langle #1 \rvert}
\newcommand\sket[1]{\lvert #1 \rangle}
\newcommand\sbraket[2]{\langle #1 \vert #2 \rangle}

\newcommand{\ind}[1]{_\mathrm{#1}}

\newcommand{\er}[1]{Eq.\,\eqref{#1}}

\usepackage{verbatim}

\usepackage{pdfpages}
\usepackage{pgffor}
\makeatletter
\AtBeginDocument{\let\LS@rot\@undefined}
\makeatother

\begin{document}


\newpage
\clearpage
\twocolumngrid

\title{Circuits as a simple platform for the emergence of hydrodynamics in deterministic chaotic many-body systems}
\author{Sun Woo P. Kim}
\email{swk34@cantab.ac.uk}
\affiliation{Department of Physics, King's College London, Strand, London WC2R 2LS, United Kingdom}
\author{Friedrich Hübner}
\affiliation{Department of Mathematics, King's College London, Strand, London WC2R 2LS, United Kingdom}
\author{Juan P. Garrahan}
\affiliation{School of Physics and Astronomy, University of Nottingham, Nottingham NG7
2RD, United Kingdom}
\affiliation{Centre for the Mathematics and Theoretical Physics of Quantum Non-Equilibrium
Systems, University of Nottingham, Nottingham NG7 2RD, United Kingdom}
\author{Benjamin Doyon}
\affiliation{Department of Mathematics, King's College London, Strand, London WC2R 2LS, United Kingdom}

\begin{abstract}
    The emergence of hydrodynamics is one of the deepest phenomena in many-body systems. Arguably, the hydrodynamic equations are also the most important tools for predicting large-scale behaviour. Understanding how such equations emerge from microscopic deterministic dynamics is a century-old problem, despite recent progress in fine-tuned integrable systems. 
     Due to the universality of hydrodynamics, the specific microscopic implementation should not matter. Here, we show that classical deterministic circuits provide a minimal, exact, and efficient platform that admits non-trivial hydrodynamic behaviour for deterministic but chaotic systems. By developing new techniques and focusing on 1D circuits as a proof of concept, we obtain the characteristic dynamics, including relaxation to Gibbs states, exact Euler equations, shocks, diffusion, and exact KPZ super-diffusion. Our methods can be easily generalised to higher dimensions or quantum circuits.
\end{abstract}

\maketitle

\begin{figure*}
    \centering
    \includegraphics[width=\linewidth]{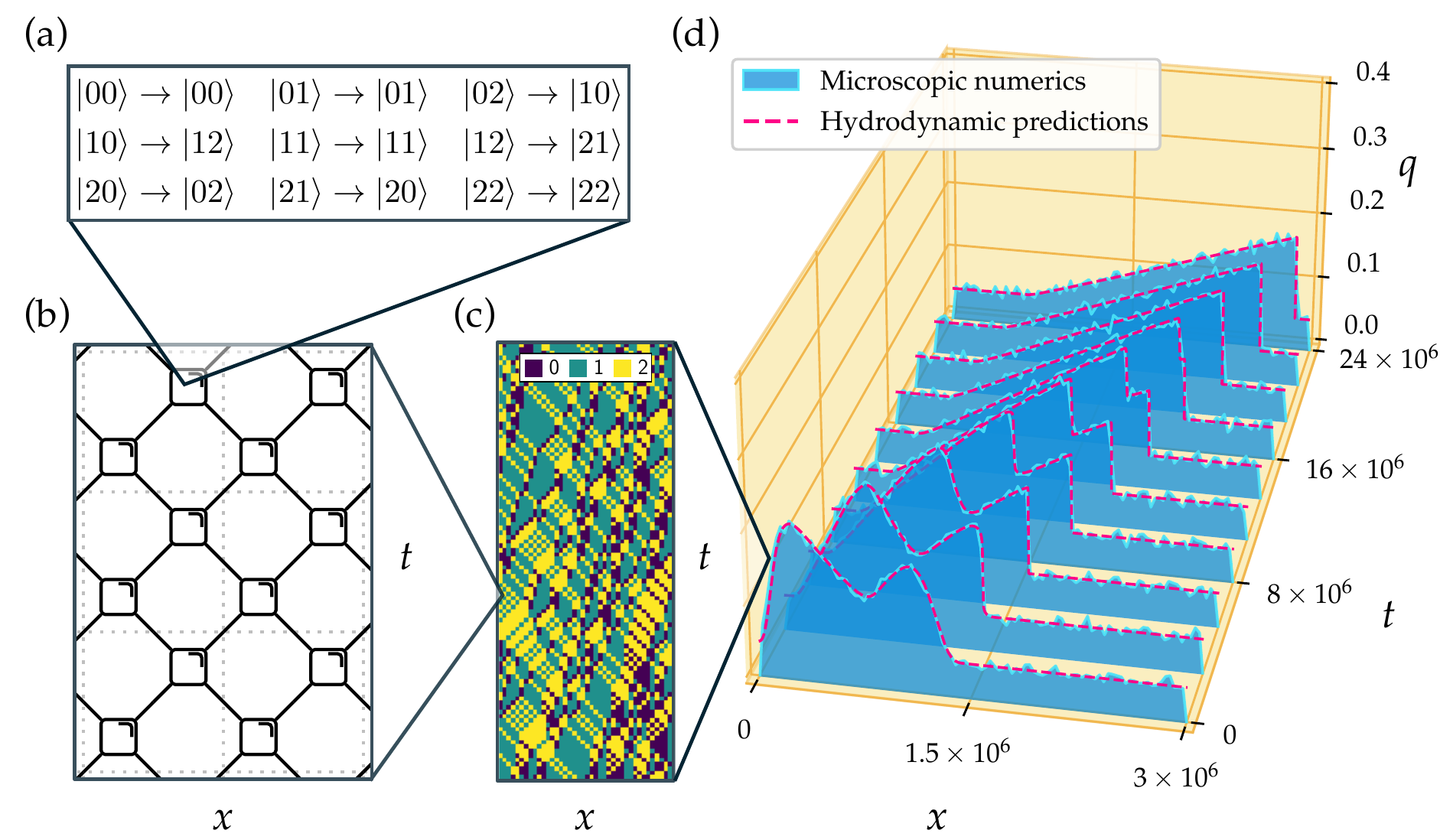}
    \caption{\textbf{Hydrodynamics of deterministic circuits,} illustrated with gate $(d, \sigma)=(3,996)$. \textbf{(a)} Local transition rules.
    \textbf{(b)} Brickwork structure of the space-time circuit.
    \textbf{(c)} Example microscopic trajectory.
    \textbf{(d)} Coarse-grained evolution of the conserved quantity, displaying formation of shocks followed by relaxation to thermal equilibrium. The results from large-scale microscopic numerics, with system size $L=3\times 10^6$ and timesteps $T=8L$ (blue shading) is accurately captured by the hydrodynamic predictions (red dashed). Here, the numerics is from a single realisation, with fluid-cell averaged over $3 \times 10^4$ sites.}
    \label{fig:main_figure}
\end{figure*}

\section{Introduction}
A longstanding approach in physics is to describe the aggregate behaviour of liquids and gases, which are particulate systems obeying microscopic laws, in terms of effective hydrodynamic equations such as Navier-Stokes~\cite{landau_fluid_mechanics}. While hydrodynamic equations are one of the best tools we have for predicting large-scale properties, understanding exactly how they emerge from microscopic interactions is a century-old problem.
Hydrodynamics~\cite{Pines1966,spohn2012large} appears to generically emerge in either classical or quantum, deterministic or stochastic systems, whenever there is (at least) one local conservation law. Crucially, the form of hydrodynamics equations and their phenomenology seem to be universal and independent of specific details of the microscopic dynamics. This suggests that one may model the microscopics in a simplified way --- an essential tool in theoretical physics --- without losing the physically relevant, emergent large-scale effects.

The key idea behind hydrodynamics is that at large scales of space and time, due to the loss of information within subtle correlations arising from an amalgamation of deterministic chaos and many-body physics, the behaviour of a system is captured by a reduced set of relevant dynamical degrees of freedom~\cite{Pines1966,spohn2012large}. Understanding how the corresponding hydrodynamic equations emerge in deterministic chaotic systems, without putting ``by hand'' phenomenological effects that induce information loss (such as noise), remains a daunting task. This is true even in highly simplified systems: analytically exact calculations are difficult, and numerical simulations are both subtle and costly in terms of time and memory requirements. This leads to the following questions: (i) What are the most basic properties of a system for it to have a non-trivial hydrodynamic scale? (ii) Can one find a class of deterministic dynamics, simple enough to allow for exact calculations, for truly large-scale simulations, and for in-depth understanding of the processes leading to hydrodynamics? (iii) Can this class of systems be complex enough to be chaotic and give rise to all non-trivial aspects of hydrodynamic phenomenology?

In this paper, we address these questions. We propose that {\em classical deterministic circuit models} form a powerful, minimal, exact, and efficient platform to study the emergence of hydrodynamics in chaotic many-body systems. Circuit models are defined in discrete space and time, with local update rules that can be deterministic or random, classical or quantum. In recent years, there has been great interest in their non-equilibrium behaviour as they combine tractability with truly many-body physics. In the quantum many-body context, random quantum circuits \cite{nahum2017quantum,nahum2018dynamics,nahum2018operator,chan2018solution,keyserlingk2018operator,fisher2023random} and dual-unitary circuits \cite{bertini2019entanglement,bertini2019exact,gopalakrishnan2019unitary}
have become central to understanding the dynamics of entanglement and thermalisation. Quantum circuits are being used to benchmark performance of quantum computers as they can be experimentally realised in existing quantum devices \cite{arute2019quantum}. In turn, classical circuit models are instances of cellular automata \cite{wolfram1983statistical}, and have been extensively studied in various contexts, including pattern formation, Turing-completeness \cite{wolfram1983statistical}, integrability \cite{bobenko1993on-two-integrable,prosen2016integrability,wilkinson2020exact,buca2021rule}, and dynamical large deviations \cite{klobas2024exact}. However, except for some specific integrable cases \cite{buca2021rule}, the intersection of classical circuits with hydrodynamics has not been explored comprehensively. 

Here, through an extensive search in the space of classical deterministic circuits made possible by the simplicity of their microscopic structure, we uncover a panoply of models with a non-trivial hydrodynamic scale. This provides a proof-of-concept of the direct connection between generic deterministic and non-trivial microscopic chaotic dynamics, and large-scale hydrodynamics. We observe a varied range of hydrodynamic phenomena, including the presence of one or more conservation laws, 
diffusion, emergence of shocks and Burgers equation~\cite{bressan2000hyperbolic}, and Kardar-Parisi-Zhang (KPZ) super-diffusion~\cite{spohn2014nonlinear}. We develop an analytical method where all conservation laws are deduced given the circuit update rule (up to some locality conditions). 
This gives us the conserved quantities admitted by the model, from which we predict the exact manifold of Gibbs states, exact thermodynamics, and exact Euler-scale hydrodynamic equations. We also develop a general numerical method --- valid for any microscopic model and efficiently realised here in classical circuits --- to visualise the manifold of Gibbs states, establishing its dimensionality and confirming relaxation onto it. By extensive simulations from long-wavelength states, we verify the emergence of the Euler-scale hydrodynamics, and confirm the predicted diffusive and KPZ super-diffusive behaviours beyond the Euler scale, obtaining the exact super-diffusion constant for the latter. 
\cref{fig:main_figure} summarises how hydrodynamics emerges: (a) the local microscopic dynamical rules, (b) implemented in a spacetime brickwork circuit, (c) give rise to chaotic trajectories, (d) for which we predict the exact large scale behaviour through hydrodynamic theory, including non-trivial phenomenology such as shocks.
While we focus on circuits in one spatial dimension for simplicity, all our methods are easy to generalise to higher dimensions or quantum circuits.

\section{Classical deterministic circuits}

In this work we study 
classical circuits with deterministic and reversible dynamics in a ``brickwork'' geometry in $1+1$ dimensions (i.e., one space and one time dimension), see \cref{fig:main_figure}. We consider configurations $\bm{a} = a_{1:L} = (a_1, \dots, a_L)$ of a one-dimensional lattice of $L$ sites with periodic boundary conditions (with $L$ assumed to be even). We use the notation ${i\!:\!j}$ to denote support on sites $\{i, i+1, \dts, j\}$, and bold symbol to denote support over the whole lattice. On each site $i$, the local configuration $a_i \in \mathcal{D}$ is one of a discrete set of $d$ states, $\mathcal{D} = \{0, 1, \cdots, d-1\}$.

Time evolution is constructed by applying at every timestep $\tau=0, 1, \dots, T-1$, where $T$ is the total time (also assumed even), local two-body gates $\mathsf{u}$ in a brickwork fashion, see \cref{fig:main_figure}(b): at step $\tau$ we apply the local gates to every other bond in the lattice, which defines the evolution operator
\begin{align}
    \bm{\mathsf{u}}_\tau := \bigotimes_{i \in \mathcal{R}_\tau} \mathsf{u}_{i:i+1},
    \label{eq:u}
\end{align}
where $\mathcal{R}_\tau$ is the set of even (resp.~odd) sites for even (resp.~odd) time $\tau$. The even and odd evolution operators \eqref{eq:u} are applied periodically every two time steps. 

A classical deterministic and reversible gate $\mathsf{u}$ are described by an invertible map $\sigma$ on two neighboring sites that maps uniquely a pair of site configurations in $\mathcal{D} \times \mathcal{D}$ to another in $\mathcal{D} \times \mathcal{D}$. Such a map can be viewed as a permutation $\sigma \in S_{d^2}$, hence there are $(d^2)!$ possible gates. We index each  gate by the lexicographic index of its corresponding permutation \footnote{
The label $\sigma$ that defines a gate is defined as follows. 
We first identify each two-site configuration $(a, a')$ in $\mathcal{D} \times \mathcal{D}$ with a number $A = da + a'$, such that $(0, 0) \leftrightarrow 0$, $(0, 1) \leftrightarrow 1$, until $(d-1, d-1) \leftrightarrow d^2-1$. Then the action of a gate corresponds to a permutation $(0, 1, \dts, d^2-1) \rightarrow A_{0:d^2-1} = (A_0, \dts, A_{d^2-1})$. The set of all permutations $\{A_{0:d^2-1}\}$ can be then indexed in lexicographic order: $\sigma=0$ corresponds to the trivial permutation $A_{0:d^2-1}=(0, 1, 2, \dts, d^2-3, d^2-2, d^2-1)$, $\sigma=1$ to $A_{0:d^2-1}=(0, 1, 2, \dts, d^2-3, d^2-1, d^2-2)$, and so on, all the way to $\sigma = (d^2)!-1$ for $A_{0:d^2-1}=(d^2-1, d^2-2, d^2-3, \dts, 2, 1, 0)$.}. 
In what follows we refer to a circuit with the notation $(d, \sigma)$ of its site dimension and the permutation code that defines the transition rule. For example, the rule of circuit 
$(d, \sigma) = (3, 996)$ is shown in \cref{fig:main_figure}(a). This model will be used to illustrate and present the general methodology in what follows.

In an abuse of notation, we will sometimes also indicate by $\sigma$ the function that defines the gate. Given the input configurations $a$ and $b$ into a gate, we write $\sigma(a,b)=(\sigma_1(a,b),\sigma_2(a,b))$ for the output two-site state, where $\sigma_{1,2}(a,b)$ are functions that define the permutation, cf.\ \cref{fig:main_figure}(a). If we denote the local state $a$ in terms of the basis vector (or ket) $|a\rangle$, then the gate 
$\mathsf{u}$ can be written in operator form 
\begin{align}
    \mathsf{u} = \sum_{a, b} |\sigma_1(a,b), \sigma_2(a, b)\rangle\langle a, b|
    \, ,
\end{align} 
or equivalently in tensor network notation,
\begin{align}
    \bra{a', b'} \mathsf{u} \ket{a, b} \; = \quad \gate{
        \squircle
        \giyeok
        \writeat{BL}{\small $a$}
        \writeat{BR}{\small $b$}
        \writeat{TL}{\small $a'$}
        \writeat{TR}{\small $b'$}
    } \quad = \;
    \delta_{a', \sigma_1(a,b)} \delta_{b', \sigma_2(a,b)}.
\end{align}
The evolution operator \eqref{eq:u} for odd times then looks
\begin{align}
    \bm{\mathsf{u}}_1 = \gates[6][1]{
        \writeat[0][0]{M}{$\cdots$}
        \squircle[1][0] \giyeok[TR][1][0] \writeat[1][0]{BLT}{$\underset{\mathrm{odd \; site}}{\uparrow}$}
        \squircle[3][0] \giyeok[TR][3][0]
        \writeat[5][0]{M}{$\cdots$}
        }
    \nonumber
    \, , 
\end{align}
\newline
and similarly for the even-time evolution operator. 

At this point the dynamics is completely deterministic for a fixed initial configuration. Since we are interested in large scale dynamics, where not all microscopic details of the system can be observed or are relevant, it is more suitable to describe the initial state of the system by a probability vector over configurations, $\lvert p_0 \rangle = \sum_{\bm{a}} p_0(\bm{a}) |\bm{a} \rangle$, which evolves over time under the circuit dynamics.

While this paper considers explicitly only $1+1$ circuits, the generalisation to the $D+1$ case is straightforward. For instance, on the hypercubic lattice of dimension $D$, the local gate would act on the $2^D$ sites at the corners of a hypercube. At even times the leftmost sites for the gates are even for each direction, and at odd times they are odd. There are $\left(d^{2^D} \right)!$ such gates for number of local states $d$.

\section{Conserved quantities (and how to find them)}
One of the most important characteristics of a dynamical system with local interactions, such as deterministic classical circuits considered here, is the set of extensive {\em conserved quantities} (CQs) that it admits. These are real functions of configurations $F(\bm{a})$ invariant under time evolution, $F(\bm{\mathsf{u}}_{n} \cdots \bm{\mathsf{u}}_1 \bm a) = F(\bm a_1)$ for some $n$, and which can be written as $F = \sum_{i} F_i$, where the {\em local density} $F_i$ depends only on the part of the configuration lying on a finite region around site $i$. $F_i$ satisfies the equivalent of a continuity equation with an associated local current in discrete space-time. Such continuity equations control the large-scale dynamics, as at large scales they become continuous conservation laws at the basis of hydrodynamics. However, one has to be careful: it has been recently understood \cite{ilievski2016quasilocal,doyon2017thermalization,doyon2022hydrodynamic,ampelogiannis2024long-time} that relaxation and hydrodynamics require the knowledge of a larger class of extensive $F$ including, but not restricted to, those with quasi-local $F_i$, although in most systems they are not expected to be present.
The search for CQs is at the basis of our results. It will be restricted to local densities. Crucially, our numerical algorithm to uncover the manifold of Gibbs states, discussed below, intrinsically accounts for {\em the complete set of CQs}.

Given a gate $(d, \sigma)$, we aim to find all CQs. For example, one could construct a CQ at the gate level by associating a local ``energy'' $f(a)$ to each state, then demand that the gate conserves it for any pair of inputs. This corresponds to the condition
\begin{align} \label{eq:gate-level-cq}
    \Big( \langle f | \langle - | + \langle - | \langle f | \Big) \mathsf{u} = \langle f | \langle - | + \langle - | \langle f |,
\end{align}
where $\bra{f} = \sum_{a} f(a) \bra{a}$ and we have introduced the ``flat'' state $\bra{-} = \sum_a \bra{a} = \begin{tikzpicture}[baseline={([yshift=-.5ex] current bounding box.center)}, scale=0.5]{\leg{L} \stub[H]{TL}}\end{tikzpicture}$
, which as the gates are permutations is always a CQ (indicating probability conservation). We now present a method to determine if there exists CQs with densities of \emph{locality} up to $(2l-1)$ for some $l$, i.e., supported on $(2l-1)$ sites.

We define the right translation matrix
\begin{align}
    \bm{\mathsf{T}} = \gates[5][1]{
        \writeat[0][0]{M}{$\cdots$}
        \leg[1][0]{D}
        \leg[2][0]{D}
        \leg[3][0]{D}
        \writeat[4][0]{M}{$\cdots$}
        }
    \, ,
\end{align}
which shifts configurations
$\cdts \lvert a \rangle_{-1} \lvert b \rangle_0 \lvert c \rangle_1 \cdts \rightarrow \cdts \lvert a \rangle_{0} \lvert b \rangle_1 \lvert c \rangle_2 \cdts$. Evolution over two steps can then be written
\begin{align}
    \bm{\mathsf{U}} := \bm{\mathsf{u}}_1 \bm{\mathsf{T}} \bm{\mathsf{u}}_1 \bm{\mathsf{T}}^{-1} = \gates[6][2]{
        \writeat[0][0]{M}{$\cdots$} \writeat[0][1]{M}{$\cdots$}
        \squircle[1][0] \giyeok[TR][1][0] \squircle[3][0] \giyeok[TR][3][0]
        \writeat[1][0]{BLT}{$\underset{\mathrm{even \; site}}{\uparrow}$}
        \squircle[2][1] \giyeok[TR][2][1] \squircle[4][1] \giyeok[TR][4][1]
        \writeat[5][0]{M}{$\cdots$} \writeat[5][1]{M}{$\cdots$}
    }.
\end{align}
Consider some $\bra{F}$ that satisfies 
$\bra{F} \bm{\mathsf{T}}^2 = \bra{F} \mu^2$ and $\bra{F} \bm{\mathsf{U}} = \bra{F} \lambda^2$ (note $[\bm{\mathsf{T}}^2, \bm{\mathsf{U}}]=0$). The latter condition can be written 
$\bra{F} (\bm{\mathsf{u}}_1 \bm{\mathsf{T}})^2 \bm{\mathsf{T}}^{-2} = \bra{F} \lambda^2$, and since $[\bm{\mathsf{u}}_1 \bm{\mathsf{T}}, \bm{\mathsf{T}}^2] = 0$, this means that $\bra{F} \bm{\mathsf{u}}_1 \bm{\mathsf{T}} = \bra{F} \lambda \mu$. 
As both $\bm{\mathsf{u}}_1$ and $\bm{\mathsf{T}}$ are permutations, $\mu$ and $\lambda$ must be phases: 
if they are roots of unity of order $m$ and $n$, respectively, then $\bra{F}$ is a CQ with spatial periodicity $m$ and temporal periodicity $n$, i.e., it is conserved after $n$ iterations of time evolution.

Now consider an ansatz for the CQ,
\begin{align}
    \langle F \rvert = \sum_{j \; \mathrm{odd}} \mu^{j} \Big( \cdts \langle - \rvert \langle f\rvert_{j:j+2l-1} \langle - \rvert \cdts \Big) =: \sum_{j \mathrm{\; odd}} \mu^j \bra{F_j}
\end{align}
with $\langle F \rvert \bm{\mathsf{T}}^2 = \langle F \rvert \mu^2$ by definition. Without loss of generality, we restrict to CQs of the type
\begin{align}
    \langle f\rvert_{1:2l} = \langle f_\mathrm{o} \rvert_{1:2l-1} \langle - \rvert_{2l} + \langle - \rvert_{1} \langle f_\mathrm{e} \rvert_{2:2l},
\end{align}
corresponding to $(2l-1)$-local CQs that are different starting on even and odd sites, and write the CQ as
\begin{align} \label{eq:conserved-quantity-ansatz}
    \langle F \rvert = \sum_{i  \; \mathrm{odd}} \mu^{i} \Big(\bra{F_i^\mathrm{o}} + \bra{F_{i+1}^\mathrm{e}}\Big),
\end{align}
where $\bra{F^{\mathrm{o},\mathrm{e}}_i} = \left(\cdts \bra{-} \bra{f_{\mathrm{o},\mathrm{e}}}_{i:i+2l-1}\bra{-}\cdts\right)$. Considering the action of $\bm{\mathsf{u}}_1$ on $\bra{F}$, we find that any $\bra{f_\mathrm{o}}$, $\bra{f_\mathrm{e}}$ that satisfies 
\begin{align}
    \begin{aligned} \label{eq:linear-depth-1}
        \Big(\langle f_\mathrm{o} \rvert \langle - \rvert + \langle - \rvert \langle f_\mathrm{e} & \rvert\Big) \mathsf{u}_{1:2l} = \Big( 
            \mu^{-1} 
            \langle f_\mathrm{e} \rvert \langle - \rvert 
            +  
            \mu
            \langle - \rvert \langle f_\mathrm{o} \rvert 
        \Big) 
        \lambda,
    \end{aligned}
\end{align}
where $\mathsf{u}_{1:2l}= \bigotimes_{i=1, \mathrm{\; odd}}^{2l-1} \mathsf{u}_{i:i+1}$, is a CQ.
\cref{eq:linear-depth-1} is a linear equation for $\bra{f_\mathrm{o}} \oplus \bra{f_\mathrm{e}}$ that can be solved to determine if there are any CQs of the form \cref{eq:conserved-quantity-ansatz}. For each gate $\mathsf{u}$, we solve \cref{eq:linear-depth-1} to determine if there are any CQs for given $(l,\mu,\lambda)$, scanning roots of unity of order $(m,n)$ for $(\mu,\lambda)$, starting from $m = n = 1$ and increasing them systematically. Further details, such as gauge fixing, are given in SM Sec. B \cite{supplemental}.

When $l=\mu=\lambda=1$ we have a \emph{simple CQ} (SCQ), that is, a single-site CQ with spatial and temporal periodicity 1. For SCQs, \cref{eq:linear-depth-1} simplifies to
\newline
\begin{align} \label{eq:single-site-linear-eq}
    \gate{\squircle \writeat{TLT}{$f_\mathrm{o}$} \stub[H]{TR}} 
    +
    \gate{\squircle \stub[H]{TL} \writeat{TRT}{$f_\mathrm{e}$}}
    =
    \gate{\leg{L} \leg{R} \writeat{TLT}{$f_\mathrm{e}$} \stub[H]{TR}} 
    +
    \gate{\leg{L} \leg{R} \stub[H]{TL} \writeat{TRT}{$f_\mathrm{o}$}}.
\end{align}
For example, solving \cref{eq:single-site-linear-eq} for $(d,\sigma)=(3,996)$, we find the SCQ
\begin{align}
    \bra{f_\mathrm{e}} = \bra{1}, \quad \bra{f_\mathrm{o}} = -\bra{0}
    \, , 
\end{align}
meaning that the number of $1$s on even sites minus the number of $0$s on odd sites is conserved. We find no other CQs up to $l, n, m \leq5$ in this circuit.

The simplest circuits one could have considered are those with $d=2$. However, an exhaustive search over all $(2^2)!=24$ gates using the methods above show that they either have no CQ, or appear to have a tower of CQs; see Sec. A of SM \cite{supplemental}. 
The first case corresponds to systems which thermalise to the infinite temperature state and have no hydrodynamics. In turn, the second case corresponds to systems likely to be integrable: these are described by the theory of generalised hydrodynamics \cite{castro2016emergent,bertini2016transport,doyon2020lecture,essler2023a-short}, which is known to have non-generic hydrodynamics (e.g., they do not admit shock formation \cite{el2011kinetic,pavlov2012generalized,hubner2024existence}) and will not be discussed here. For these reasons we will consider circuits with local dimension $d=3$, focusing on gates with a non-zero but finite number of CQs in order to study the full range of hydrodynamic phenomenology.

\begin{figure*}[t]
    \centering
\includegraphics[width=\linewidth]{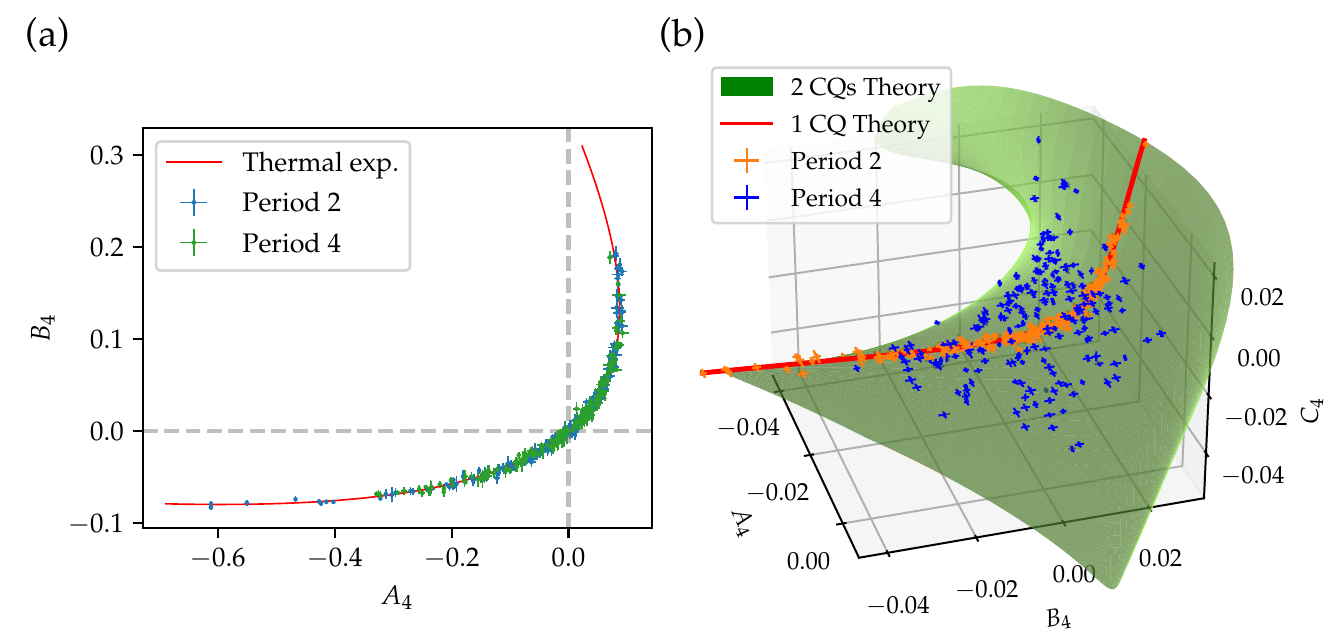}
    \caption{\textbf{Expectation values of observables embedded in low-dimensional manifolds.} Randomly generated observables of random initial states overlaid on theoretical thermal expectation values, for $\beta \in [-5, 5]$ for $L=20000$, $T=1000$, averaged over the last $10$ even timesteps. \textbf{(a)} For gate $(d, \sigma)=(3, 996)$, we expect only a single CQ. For both kinds of initial states with spatial periodicities $m=2,4$, we see good evidence that the observables lie on a one-dimensional manifold, and that their values agree with the thermal expectations. \textbf{(b)} For gate $(d, \sigma) = (3, 229117)$, we expect two conserved quantities, one with $m_1 = 2$ and another ``staggered'' quantity with $m_2=4$. For initial states with $m=2$, the data lies on a one-dimensional curve corresponding to thermal expectation on only the first CQ, as it does not couple to the staggered CQ. Using instead initial states with $m=4$, we find that data lies on a two-dimensional surface of thermal expectation on both CQs.} \label{fig:nikes}
\end{figure*}

\section{Manifold of stationary states} \label{sec:verifying}

With the previous method, we can exactly determine CQs for given locality and periodicities. While the above cannot rule out CQs with higher localities and periodicities or CQs in the larger class of extensive observables referred to above, we can independently determine their number by studying stationary states of the dynamics. 
According to general principles of many-body physics, we expect that any initial state will relax in such a way that averages of local observables be given by ``thermal states" (maximal entropy states) with Gibbs probability 
\begin{align} \label{eq:beta-state}
    \lvert p_{\vec{\beta}} \rangle 
    = 
    \frac1{\mathcal{Z}(\vec{\beta})}
    \,
    \sum_{\bm{a}} 
        e^{- \vec{\beta} \cdot \vec{F}(\bm{a})} 
        \ket{\bm{a}}
        \, .
\end{align}
Here, $\vec{F}(\bm{a}) = (\sbraket{F^{(\alpha)}}{\bm{a}})_{\alpha=1:N_{\mathrm{cq}}}$ is the set of CQs, $\vec\beta = (\beta_\alpha)_{\alpha=1:N_{\mathrm{cq}}}$ is the set of conjugate parameters (or Lagrange multipliers), $N_{\mathrm{cq}}$ is the total number of CQs, and we use the arrow notation to indicate a vector in CQ-species space. Since \er{eq:beta-state} only has $N_{\mathrm{cq}}$ degrees of freedom, if we plot the thermalised values of $M > N_{\mathrm{cq}}$ observables starting from different initial conditions, we should find that they lie on $N_{\mathrm{cq}}$-dimensional submanifold. If that is the case, then we have the full set of CQs, and in particular, we rule out the existence of CQs with quasi-local density. 

The simplest non-trivial situation is for a circuit which has one SCQ. The dynamics is generated then by what we call a \emph{simple gate} (SG). For SGs, the thermal state factorises into a product state, $\sket{p_\beta} = \cdts \sket{p_\beta^\mathrm{e}} \sket{p_{\beta}^\mathrm{o}} \cdts$, where $\sket{p^{\mathrm{e},\mathrm{o}}_{\beta}} = (1/z_\beta^{\mathrm{e},\mathrm{o}})\sum_a \exp[-\beta f^{\mathrm{e},\mathrm{o}}(a)] \ket{a}$, with $z^{\mathrm{e},\mathrm{o}}_\beta = \sum_a \exp[-\beta f^{\mathrm{e},\mathrm{o}}(a)]$. The action of SGs on their thermal states is
\begin{align}
    \gate{\squircle \writeat{BLT}{$p^\mathrm{e}_\beta$} \writeat{BRT}{$p^\mathrm{o}_\beta$}} = \gate{\leg{L} \leg{R} \writeat{BLT}{$p^\mathrm{o}_\beta$} \writeat{BRT}{$p^\mathrm{e}_\beta$}}. \\
    \nonumber
\end{align}
In general, for CQs with larger support, \cref{eq:beta-state} corresponds to a matrix product state (MPS).

To detect if there is zero, one, or more CQs, we proceed as follows: (i) we choose two random quantities $A_l$, $B_l$, supported on $l$ sites and orthogonal to the flat state and each other (i.e. $\braket{A_l}{B_l} = 0$); (ii) we construct an initial many-body probability distribution with spatial periodicity $m$ as the product of a random $m$-body probability $p(a_{1:m})$; (iii) we sample an initial configuration from this initial distribution; (iv) we then run the dynamics for $L \gg T \gg t_{\mathrm{thermo}}$ such that boundary conditions are not seen but the system relaxes.

We obtain $t_\mathrm{thermo}$ by plotting observables with respect to time and locating when they plateau. We estimate the thermal averages $\expval{A_{l}}_{\vec{\beta}}$ and $\expval{B_{l}}_{\vec{\beta}}$, where 
\begin{equation}
    \label{eq:thav}
    \expval{X}_{\vec{\beta}} := 
        \sum_{\bm{a}} 
        X(\bm{a}) \,
        e^{- \vec{\beta} \cdot \vec{F}(\bm{a})} 
        / \mathcal{Z}(\vec{\beta})
        =
        \langle X | p_{\vec{\beta}} \rangle 
    \, ,
\end{equation}
as running averages over $n$ (alternate) timesteps after thermalisation. By parametrically ploting two such numerical estimates over a range of initial states we can infer $N_{\mathrm{cq}}$. If there are no CQs, the plot should converge to a single point corresponding to the single maximum entropy state. 
If $N_{\mathrm{cq}} =1$, the plot will converge to a one-dimensional curve on the plane, while for $N_{\mathrm{cq}} > 1$, it will become two-dimensional. In \cref{fig:nikes}(a) we show such plot for the circuit $(d, \sigma)=(3, 996)$, for $l=4$, $m=2,4$, and $n=2$. We see that the numerical averages converge on a one-dimensional manifold indicating only a single CQ. The numerics agrees well the theory, confirming that $(3, 996)$ is a SG.

The generalisation to $N_{\mathrm{cq}}>1$ is straightforward. For example, if two CQs are expected, then expectations of three randomly chosen observables should occupy a two-dimensional embedded. We show such case in \cref{fig:nikes}(b) for circuit $(d, \sigma) = (3, 229117)$. In higher dimensions, where visualisation becomes difficult, box-counting dimensions \cite{barabasi1995fractal} could be used to determine the dimension of the sub-manifold and thus confirm $N_{\mathrm{cq}}$. 

\section{Constructing the hydrodynamic equations}
After the CQs are verified, we construct the microscopic continuity equations~\cite{spohn2012large,doyon2020lecture}. For the $N_{\mathrm{cq}}$ CQs, with spatial periodicities $\{m_\alpha\}_{1:N_{\mathrm{cq}}}$, we define a supercell of size $m = \mathrm{lcm}(\{m_\alpha\}_{1:N_{\mathrm{cq}}}, 2)$, and charge in the supercell
\begin{align} 
    \label{eq:dressed-charge}
    \sbra{q^{(\alpha)}_i} 
    = 
    \frac{1}{m}\sum_{j \mathrm{\; even}}^{m-2} \mu_\alpha^{j} \sbra{F^{(\alpha)}_{i + j}}
    \, .
\end{align}
Then, given temporal periodicities $\{n_\alpha\}_{1:N_{\mathrm{cq}}}$, we consider the change in charge in this supercell after a superunit of time $n = \mathrm{lcm}(\{n_\alpha\}_{1:N_{\mathrm{cq}}}, 2)$. This leads to the microscopic continuity equation
\begin{align} \label{eq:microscopic-continuity}
    \frac{1}{n}\left(\sbra{q^{(\alpha)}_{i, n}} - \sbra{q^{(\alpha)}_{i, 0}}\right) & = - \frac{1}{m}\left(\sbra{\j_{i+m, 0}^{(\alpha)}} - \sbra{\j_{i, 0}^{(\alpha)}}\right).
\end{align}
Here, $\sbra{q^{(\alpha)}_{i,\tau}} = \sbra{q^{(\alpha)}_{i}} {\bm{\mathsf{U}}}^{\tau/2}$ is the time-evolved charge, $\sbra{\j^{(\alpha)}_i}$ is the current, and $\sbra{\j^{(\alpha)}_{i+m, 0}} = \sbra{\j^{(\alpha)}_{i, 0}} \bm{\mathsf{T}}^{-m}$ is the current shifted by $m$ sites. For a SCQ
\newline
\begin{align}
    \langle \j^{(\alpha)}_i \rvert = \cdots \inlinetikz{\leg{L} \stub[H]{TL}} \left(
        - \gate{
        \squircle \writeat{BRT}{$\underset{i}{\uparrow}$}
        \writeat{TLT}{$f_\mathrm{o}$} \stub[H]{TR}
        }
        +
        \gate{
            \leg{L} \leg{R} \writeat{TRT}{$f_\mathrm{e}$} \stub[H]{TL} \writeat{BRT}{$\underset{i}{\uparrow}$}
        } 
    \right) \inlinetikz{\leg{L} \stub[H]{TL}} \cdots. \\
    \nonumber
\end{align}
The general case is detailed in SM Sec. C \cite{supplemental}.
 
Next, we derive the macroscopic hydrodynamic equations. These are PDEs in continuous variables $x=i$ and $t=\tau$, which interpolate between space-time lattice of the circuit. If $L$ is very large, it is natural to assume that the state is in local equilibrium, with $\vec \beta(x)$ varying on the large scale $L$. Then, the charges and currents are given by the spatially-varying averages, 
$\vec{q}(x) := \expval{\vec{q}}_{\vec{\beta}(x)}$ and $\vec{\j}(x) := \expval{\vec{\j}}_{\vec{\beta}(x)}$, 
cf. \cref{eq:thav}. By inverting the first relation, we can write the currents as functions of the charges and deduce the Euler-scale hydrodynamics equation
\begin{equation}
    \label{eq:exact-continuity-eq}
    \partial_t \vec{q}(x) 
    = 
    -\partial_x \vec{\j}(x) 
    = 
    - \vecev{V} \partial_x \vec q(x)
    \, , 
\end{equation}
where $\big[\vecev{V}\big]_{\alpha, \alpha'} = \partial \j^{(\alpha)} / \partial q^{(\alpha')}$ is the ``flux Jacobian".

For SGs, we have that (see SM Sec. C \cite{supplemental}) 
\begin{align} 
    \label{eq:simple-current}
    \expval{\j}_\beta = \expval{f}_{\mathrm{o},\beta} - \expval{f}_{\mathrm{e},\beta},
\end{align}
i.e.\ the average current is zero when the CQ is the same on even and odd sites. Therefore, the  construction \cref{eq:gate-level-cq} does not allow for non-zero currents.

\begin{figure}
    \centering
    \includegraphics[width=0.9\linewidth]{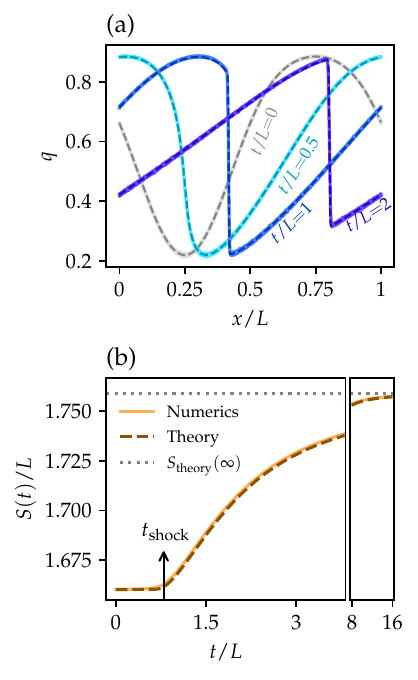}
    \caption{\textbf{Rapid thermalisation from formation of shocks in deterministic circuits,} illustrated with gate $(d, \sigma)=(3, 996)$. \textbf{(a)} Formation of shocks. Microscopic numerics for $L=10^5$ averaged over $10^5$ samples (solid lines, linewidth corresponds to standard deviation) verifies weak hydrodynamic solutions (dashed lines) given by Rankine-Hugoniot conditions. \textbf{(b)} Entropy production from shocks. $S(t)$ is constant until shock formation time $t_{\mathrm{shock}} \approx 0.95L$, which facilitates entropy production until saturating to a theoretical maximum value $S_\rm{theory}(\infty) \approx 1.758L$ at $t_{\rm{eq}} \approx 16 L$. Here, $L=8\times 10^4$ and the result is averaged over $10^3$ samples.} \label{fig:996_hydro}
\end{figure}

For the circuit $(d,\sigma)=(3, 996)$, we find that the average charge and current are
\begin{align}
    \expval{q}_{\beta} = \frac{1}{2 e^{\beta}+1}+\frac{2}{e^{\beta}+2}-1,
\end{align}
\begin{align}
    \expval{\j}_{\beta} = \frac{1}{-2 e^{\beta }-1}+\frac{2}{e^{\beta}+2}-1,
\end{align}
resulting in $\j(q) = \frac{1}{3} \left(2 - \sqrt{9 q^2+16}\right)$ and a simple Euler-scale hydrodynamic equation,
\begin{align} \label{eq:996-hydro-eq}
    \partial_t q(x) = -v(q(x)) \partial_x q(x),
\end{align}
where the velocity is given by $v(q) = {3q}/{\sqrt{16+9q^2}}$. 
In \cref{fig:996_hydro}(a), we show excellent agreement between \cref{eq:996-hydro-eq} and the microscopic numerics for the circuit. Here, the evolution is deterministic. To obtain observables from distributions over initial states, we average observables over multiple samples. Alternatively, a single sample can be coarse-grained, see \cref{fig:main_figure}(d). This demonstrates that hydrodynamics is self-averaging.

With the change of variables $\rho = v(q)$, \cref{eq:996-hydro-eq} can be mapped to the Burgers equation for $\rho$. Note that it is impossible to obtain Burgers equation as the hydrodynamics of time-continuous Hamiltonian 
systems~\footnote{
    In 1D Hamiltonian systems, the Hamiltonian is always conserved, and the energy current in a Gibbs state vanishes~\cite{kobayashi2022vanishing}. Hence, if only the Hamiltonian is conserved, the Euler hydrodynamic equations are trivial $\partial_t q = 0$. The only 1D Hamiltonian systems that have a non-trivial hydrodynamics are those with multiple CQs (implying that the hydrodynamic equations are multiple coupled equations).
} (circuits have is no conserved Hamiltonian as they break time-translation symmetry). To the best of our knowledge, the circuits here are the first examples of deterministic and closed microscopic models that give rise to the Burgers equation. This demonstrates that obtaining hydrodynamic phenomenology from circuits can be simpler than Hamiltonian systems, making circuits an ideal platform to study hydrodynamics.

\begin{figure*}[t]
    \centering
    \includegraphics[width=\linewidth]{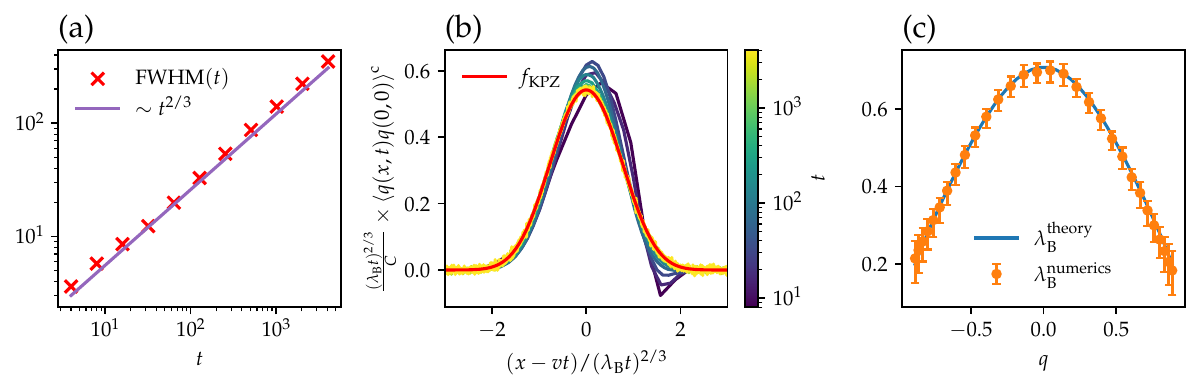}
    \caption{\textbf{KPZ super-diffusion in deterministic circuits with one conserved quantity,} illustrated with gate $(d, \sigma)=(3, 996)$. (a) Super-diffusive scaling of the full-width half maximum (FWHM) of the correlation function $\langle q(x,t) q(0,0) \rangle^\rm{c}_{\beta=0}$ obtained from microscopic numerics (red crosses, $L=2^{14}$ and averaged over $10^6$ samples), consistent with $t^{2/3}$ scaling (purple line). (b) Data collapse of $\langle q(x,t) q(0,0) \rangle^\rm{c}_{\beta=0}$, which also converges to the universal KPZ function $f_\rm{KPZ}$ (red line). (c) Theoretical prediction of the super-diffusive constant $\lambda_\rm{B}$ (blue line) compared to the value extracted from the numerics (orange markers, $L=10^3$ and averaged over $10^6$ samples).} \label{fig:996-kpz}
\end{figure*}

\begin{figure*}
    \centering
    \includegraphics[width=\linewidth]{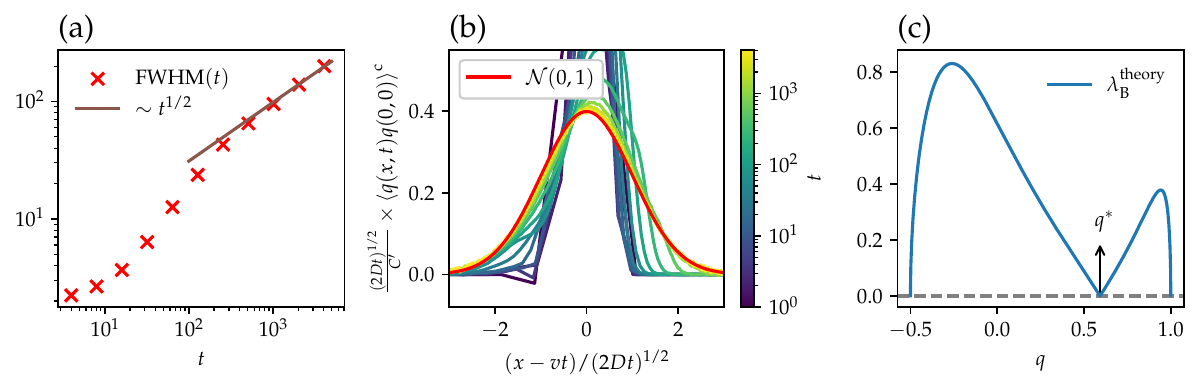}
    \caption{\textbf{Coexistence of diffusion and KPZ superdiffusion in deterministic circuits}, illustrated with gate $(d,\sigma)=(3,1092)$. At special point $q^*=\sqrt{3}\sin(\pi/9)$, \textbf{(a)} the correlations show diffusive $t^{1/2}$ scaling, and \textbf{(b)} converge to a standard normal $\mathcal{N}(0,1)$ when suitably scaled with the diffusion, $D=0.91690(1)$, and height, $C'=0.17840(1)$, constants. \textbf{(c)} Away from $q^*$, the correlations show superdiffusion.} \label{fig:1092_diffusion}
\end{figure*}

\section{Shocks, entropy and thermalisation}
When discontinuities are absent, Euler-scale hydrodynamic equations conserve the total entropy of the system $S=\int\dd{x}s(q(x))$, with a local conservation law for the entropy density $s(q(x))$, see \cite{doyon2020lecture}. But typically, Euler-scale equations, which are examples of hyperbolic PDEs, develop shocks after finite time \cite{bressan2000hyperbolic}. This is illustrated \cref{fig:main_figure}(d) for the hydrodynamic equation \eqref{eq:996-hydro-eq} of gate $(d,\sigma) = (3,996)$. Shocks appear as discontinuities in profiles of densities, where entropy is not conserved. Thus one looks for ``weak solutions'' of the hydrodynamic equations, where the continuity equation \eqref{eq:exact-continuity-eq} is written in integral form. Weak solutions are not unique, and additional conditions must be imposed, such as non-negative entropy production at shocks. Crucially, entropy production then guarantees that thermalisation may occur in finite Euler-scale time.
The concave entropy density $s(q)$ is given by
\begin{align} \label{eq:concave-entropy-density}
    s(q) & := \expval{-\ln p}_{\beta(q)}.
\end{align}
One may evaluate the shock formation time
$t_\rm{shock}$ as
\begin{align} \label{eq:t-shock}
    t_\rm{shock} = \left(\sup_x v'(q_0(x))\partial_x q_0(x) \right)^{-1},
\end{align}
where $q_0(x)$ is the initial condition for the charge profile. 
The velocity of shocks $v_\rm{shock}$ can be predicted by imposing charge conservation. This is known as the Rankine–Hugoniot condition~\cite{bressan2000hyperbolic},
\begin{align} \label{eq:rankine-hugoniot}
    v\ind{shock} = \Delta \j/\Delta q,
\end{align}
where for any function $h(x)$, $\Delta h = h(x\ind{shock}^+) - h(x\ind{shock}^-)$ is its jump across the shock. A single shock produces entropy
\begin{align}
    \dv{t} S &= \Delta g - v\ind{shock} \Delta s  \geq 0,
\end{align}
which is always positive if $v'(q) \geq 0$, and where $g(q)$ is the entropy current, with $g'(q) = s'(q)v(q)$. Once the shock appears, entropy increases until equilibration: global relaxation from inhomogeneous initial condition occurs on Euler timescale. Note that this is markedly different from integrable systems and more generally ``linearly degenerate" systems, where the absence of shocks implies entropy conservation on the Euler scale resulting in significantly longer thermalisation timescales. 

The above predictions can be explicitly verified for classical deterministic circuits. We show this in \cref{fig:996_hydro} for circuit $(d, \sigma) = (3, 996)$: in panel (a) the charge profiles from hydrodynamic predictions of \cref{eq:t-shock,eq:rankine-hugoniot} (dashed lines) are compared with those from the microscopic dynamics (solid lines) for an initial profile $q_0(x)$ (grey line). Note that hydrodynamics is still accurate even after shock formation. Moreover, $s(q)$ can be explicitly evaluated as
\begin{align}
    s(q) = q\ln \tfrac{5q+\sqrt{16+9q^2}}{4(1+q)} + \ln\tfrac{16-9q+5\sqrt{16+9q^2}}{(1+q)(-5+\sqrt{16+9q^2})}.
\end{align}
Combining with hydrodynamic predictions, we compare entropy predictions from hydrodynamics with microscopic numerics in \cref{fig:996_hydro}(b), which shows excellent agreement.

\section{Kardar-Parisi-Zhang superdiffusion from non-linear fluctuating hydrodynamics}
The hydrodynamic description enables us not only to describe one-point averages, but also higher-order correlation functions. For instance, for the two-point function, we can predict its KPZ scaling, its exact asymptotic shape and its superdiffusion constant $\lambda\ind{B}$ using the non-linear fluctuating hydrodynamic theory of Ref. \cite{spohn2014nonlinear}. Here, we describe the procedure for SGs, though generalisation to more and other kinds of CQs is straightforward. Expanding \cref{eq:exact-continuity-eq} around a constant background charge, $q(x, t) = q_0 + \delta q(x, t)$, where $q_0 = \expval{q}_\beta$, we have
\begin{align} \label{eq:hydro-expansion}
    \partial_t \delta q = - \partial_x \left( v \delta q + \frac{1}{2} \j''(q_0) \delta q^2 + \cdots \right),
\end{align}
where $v = v(q_0)$.
If $\j''(q_0) \neq 0$, we predict that the correlation function $\expval{q(x, t) q(0, 0)}_\beta^\mathrm{c}$ will be KPZ superdiffusive, i.e.
\begin{align} \label{eq:kpz-correlations}
    \expval{q(x, t) q(0, 0)}_\beta^\mathrm{c} = \frac{C}{(\lambda_\mathrm{B} t)^{2/3}} f_\mathrm{KPZ} \left( \frac{x - v t}{(\lambda_\mathrm{B} t)^{2/3}} \right).
\end{align}
Here $C = \int dx \expval{q(x, 0)q(0, 0)}_{\beta}^\mathrm{c} = \tfrac{1}{2} \, \mathrm{var}[q_{1:2}]_\beta $, $f_\mathrm{KPZ}$ is the universal KPZ scaling function, and the superdiffusion constant is (see Sec. E
of SM for details \cite{supplemental})
\begin{align}
    \lambda_\mathrm{B} = \sqrt{2 C} \abs{\j''(q_0)}.
\end{align}
\cref{fig:996-kpz} shows that these theoretical predictions are consistent with the microscopic simulations.

Note that gates with $\abs{\j''(q^*)}=0$ for some $q^*$ can support both diffusive and superdiffusion correlations. At $q^*$ the superdiffusion constant goes to zero and we predict diffusive correlations. For example, the SG $(d,\sigma)=(3, 1092)$, with rules
\begin{align}
    \begin{aligned}
        &\lvert00\rangle\rightarrow\lvert00\rangle\quad \lvert01\rangle\rightarrow\lvert10\rangle\quad \lvert02\rangle\rightarrow\lvert20\rangle\\
        &\lvert10\rangle\rightarrow\lvert01\rangle\quad \lvert11\rangle\rightarrow\lvert11\rangle\quad \lvert12\rangle\rightarrow\lvert12\rangle\\
        &\lvert20\rangle\rightarrow\lvert02\rangle\quad \lvert21\rangle\rightarrow\lvert21\rangle\quad \lvert22\rangle\rightarrow\lvert22\rangle,
    \end{aligned}
\end{align}
and SCQ $\bra{f_\rm{e}} = -\frac{1}{2} \bra{0}$, $\bra{f_\rm{o}} = \frac{1}{2} \bra{1} + \bra{2}$, has diffusive correlations at $q^*=\sqrt{3} \sin(\pi/9)$, and superdiffusive correlations elsewhere, see \cref{fig:1092_diffusion}. More precisely, if $\abs{q-q^*} \sim \delta$ is small but finite, then diffusive phenomenology will only be a transient effect until times of order $\sim 1/\delta^4$~\footnote{This simple estimate is obtained by equating the diffusive width $\sim t^{1/2}$ with the super-diffusive width $(\lambda_\rm{B} t)^{2/3} \sim (\delta \times t)^{2/3}$.}, after which it is overtaken by KPZ superdiffusion.

\section{Other Hydrodynamic phenomenology}
Our search of $d = 3$ gates reveals a wealth of other hydrodynamic
phenomenology. We discuss representative
gates of some of them.

\noindent $\bullet$ {\bf Staggered conserved quantities.}
There are circuits with CQs with nontrivial periodicities $m, n > 1$. An example is $(d, \sigma) = (3, 229117)$, with rules 
\begin{align}
    \begin{aligned}
        & \lvert00\rangle\rightarrow\lvert12\rangle\quad \lvert01\rangle\rightarrow\lvert22\rangle\quad \lvert02\rangle\rightarrow\lvert20\rangle\quad \\
        & \lvert10\rangle\rightarrow\lvert02\rangle\quad \lvert11\rangle\rightarrow\lvert00\rangle\quad \lvert12\rangle\rightarrow\lvert10\rangle\quad \\
        & \lvert20\rangle\rightarrow\lvert01\rangle\quad \lvert21\rangle\rightarrow\lvert11\rangle\quad \lvert22\rangle\rightarrow\lvert21\rangle
        \, .
    \end{aligned}
\end{align}
Here, there is one SCQ,
\begin{align}
    \sbra{f^{(1)}_\mathrm{e}} 
    = 
    \bra{1} + \bra{2}, 
    \quad 
    \sbra{f^{(1)}_\mathrm{o}} 
    = \bra{0} + 2 \bra{1} + \bra{2}
    \, ,
\end{align}
and a second CQ with $\mu_2=\lambda_2=e^{i\pi/2}$,
\begin{align}
    \label{eq:staggered}
    \sbra{f^{(2)}_\mathrm{e}} 
    = 
    \bra{0} - \bra{1} + \bra{2}, 
    \quad \sbra{f^{(2)}_\mathrm{o}} 
    = \bra{1} + 2 \bra{2}
    \, ,
\end{align}
meaning it is conserved only after $n_2=4$ timesteps and is periodic over $m_2=4$ translations. To confirm that there are only two CQs in this circuit, we extend the method above by choosing three random observables, $(A_4,B_4,C_4)$, and plot their thermal expectation values: as shown in \cref{fig:nikes}(b), these values lie on a two-dimensional surface as predicted by theory, \er{eq:beta-state}. Since \er{eq:staggered} changes sign every two sites, if the initial state has period-2, the hydrodynamics decouples~\footnote{
    In general, the hydrodynamics of a SCQ and a staggered CQ is coupled. However, it can be shown that if the initial state is infinite temperature, the staggered CQ remains as if in that state and decouples from non-staggered CQs, see Sec. D
    in SM \cite{supplemental} for details.
} and the data falls on a one-dimensional curve, corresponding to the $\beta_2=0$ predictions. In comparison, for circuit $(d,\sigma)=(3, 996)$, both period-2 and period-4 initial states lie on a one-dimensional curve, see \cref{fig:nikes}(a).

\smallskip

\begin{figure}
    \centering
    \includegraphics[width=0.85\linewidth]{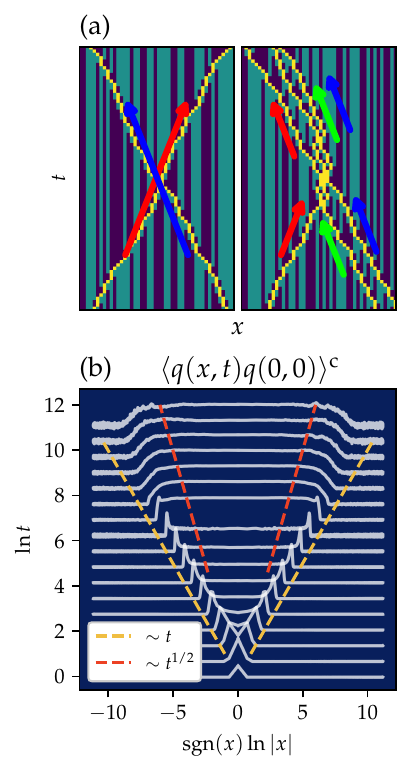}
    \caption{\textbf{FPUT phenomenology in deterministic circuits},  illustrated with gate $(d, \sigma)=(3, 2312)$. \textbf{(a)} Trajectories of collisions between $\ket{2}$s in a bath of $\ket{0}$s and $\ket{1}$s. For two-particle collisions, the $\ket{2}$s conserve momentum (left). For three-or-more particle collisions, momentum is not conserved in general (right). \textbf{(b)} Due to approximate integrability at low fillings, the correlations (white lines) show ballistic $\sim t$ spreading at early times (dashed yellow lines). Eventually, the three-or-more particle collisions facilitate diffusive $\sim t^{1/2}$ correlations (dashed red lines). Here, $\beta = -1.5$, and simulations were carried out with $L=2^{16}$ and $T=2^{14}$ and averaged over $10^4$ samples.} \label{fig:2312_fput}
\end{figure}

\noindent
$\bullet$ {\bf Fermi–Pasta–Ulam–Tsingou (FPUT) phenomenology.} The circuit $(d,\sigma)=(3, 2312)$ has a SG with rules
\begin{align}
    \begin{aligned}
        & \lvert00\rangle\rightarrow\lvert00\rangle\quad \lvert01\rangle\rightarrow\lvert01\rangle\quad \lvert02\rangle\rightarrow\lvert12\rangle\quad \\
        & \lvert10\rangle\rightarrow\lvert10\rangle\quad \lvert11\rangle\rightarrow\lvert11\rangle\quad \lvert12\rangle\rightarrow\lvert20\rangle\quad \\
        & \lvert20\rangle\rightarrow\lvert21\rangle\quad \lvert21\rangle\rightarrow\lvert02\rangle\quad \lvert22\rangle\rightarrow\lvert22\rangle.
    \end{aligned}
\end{align}
The SCQ simply counts number of $\ket{2}$s in even and odd sites, $\bra{f_{\rm{e}, \rm{o}}} = \bra{2}$. From \cref{eq:simple-current} we have $\j(q)=\j''(q)=0$, predicting diffusive correlations. However, at low densities, there is a transient state that is characteristic of integrability, precluding this diffusive effect, in a way that is similar to the FPUT phenomenology typically observed in one-dimensional systems with energy and momentum conservation~\cite{10.1063/1.1855036}. Indeed, looking at trajectories of $\ket{2}$'s, we observe that they propagate linearly, including at two-body collisions where they pass through each other, as shown on the left of \cref{fig:2312_fput}(a) (see SM Sec. F for details \cite{supplemental}). 
This is similar to what happens by energy and momentum conservation in Hamiltonian particle systems. Only when three or more $\ket{2}$'s meet is there a non-trivial interaction that breaks the linear trajectories, and eventually leads to diffusion, as shown on the right of \cref{fig:2312_fput}(a). Therefore, at low density of $\ket{2}$'s and short timescales, all trajectories are ballistic as in integrable systems, only becoming diffusive at later times, see \cref{fig:2312_fput}(b).

\smallskip

\noindent
$\bullet$ {\bf Anomalous current fluctuations.} Shocks and KPZ superdiffusion are observed in generic hydrodynamic equations. However, in certain systems satisfying linear degeneracy conditions
\footnote{
    Intuitively speaking, linear degeneracy means that hydrodynamic modes do not self-interact.
}, 
shock formation is absent \cite{bressan2000hyperbolic}. This includes models with vanishing currents, such as circuit $(d,\sigma)=(3, 2312)$ discussed previously, but also integrable systems. In linear degenerate systems, corrections to Euler hydrodynamics are typically expected to be diffusive \cite{hubner2024diffusive}. However, more nuanced universal behavior can emerge. Recently, there has been interest in anomalous diffusion due to anomalous current fluctuations. This behavior has also been observed classical brickwork circuits in Refs. \cite{krajnik2022exact,yoshimura2024anomalous}, which correspond to $(d,\sigma)=(3, 12990)$ in our notation.

\section{Outlook}
In this paper we introduced powerful but straightforward methods to deduce the conserved quantities of classical deterministic brickwork circuits and their exact hydrodynamic equations. Given the efficiency of performing microscopic simulations in such circuits, we convincingly verified the hydrodynamic predictions. 
As an important example, we described the behaviour of circuit $(d, \sigma)= (3,996)$, to our knowledge the first case of a deterministic, closed system with a single conserved quantity whose microscopics yield at large scales the Burgers equation and KPZ superdiffusion correlations. We stress the simplicity and generality of our setting here, compared to the complications in the very few other deterministic interacting models (integrable systems and certain classical 1D chains) where such formulations are possible \cite{spohn2012large}.
One reason for this simplicity is the breaking of time-translation invariance in our circuits. Looking forward, this circuit approach should help understand all aspects of hydrodynamic theories to finally prove their emergence from microscopics.

The techniques we developed here readily extend to higher-dimensional and to quantum circuits. In the quantum case, the continuum of possible two-body gates would allow to mimic the behavior of Hamiltonian systems with a smoothly-varying external potential. Hydrodynamic theories for quantum circuits are particularly interesting as they provide a way to benchmark the performance of quantum computers at large scale~\cite{rosenberg2024dynamics}. Since hydrodynamic evolution in $D>1$ generically displays turbulence rather than shocks, an important avenue for research is to find simple circuit models that can provide insight into the poorly understood physics of turbulence~\cite{landau_fluid_mechanics}.

\section{Acknowledgements}
SWPK is supported by Engineering and Physical Sciences Research Council (EPSRC) DTP
International Studentship Grant Ref.\ No.\ EP/W524475/1. 
FH is funded by the Faculty of Natural, Mathematical \& Engineering Sciences at King's College London. 
JPG is supported by EPSRC under grant EP/V031201/1 and by the Leverhulme Trust RPG-2024-112.
BD is supported by EPSRC under grants EP/W010194/1 and EP/Z534304/1. 
Numerical computations were partially done in Julia~\cite{bezanson2017julia}, in particular using the \texttt{ApproxFun.jl}~\cite{ApproxFun.jl-2014} and \texttt{Hecke}~\cite{nemo} packages, using the CREATE cluster~\cite{kings}.

\bibliographystyle{apsrev4-2} 
\bibliography{bibliography} 



\section*{Supplementary material}

# Supplementary Material for "Circuits as a simple platform for the emergence of hydrodynamics in deterministic chaotic many-body systems"

Sun Woo P. Kim,[1, *] Friedrich Hübner,[2] Juan P. Garrahan,[3,4] and Benjamin Doyon[2]

[1]*Department of Physics, King's College London, Strand, London WC2R 2LS, United Kingdom*
[2]*Department of Mathematics, King's College London, Strand, London WC2R 2LS, United Kingdom*
[3]*School of Physics and Astronomy, University of Nottingham, Nottingham NG7 2RD, United Kingdom*
[4]*Centre for the Mathematics and Theoretical Physics of Quantum Non-Equilibrium Systems,
University of Nottingham, Nottingham NG7 2RD, United Kingdom*

## A. GATES FOR $d = 2$

| $\sigma$ | Name | Rules | # CQs with support $\leq l$ | | | | | | | | | | | (Likely) classification |
|---|---|---|---|---|---|---|---|---|---|---|---|---|---|
| | | | Normal | | | | | Staggered | | | | | |
| | | | 1 | 2 | 3 | 4 | 5 | 1 | 2 | 3 | 4 | 5 | |
| 0 | Identity | $00 \to 00, 01 \to 01, 10 \to 10, 11 \to 11$ | 3 | 9 | 33 | 129 | 513 | 13 | 49 | 193 | 769 | 3073 | Integrable |
| 1 | West | $00 \to 00, 01 \to 01, 10 \to 11, 11 \to 10$ | 1 | 1 | 1 | 1 | 1 | 1 | 1 | 1 | 1 | 1 | Fully chaotic |
| 2 | SWAP | $00 \to 00, 01 \to 10, 10 \to 01, 11 \to 11$ | 3 | 5 | 9 | 17 | 33 | 13 | 25 | 49 | 97 | 193 | Integrable |
| 3 | | $00 \to 00, 01 \to 10, 10 \to 11, 11 \to 01$ | 2 | 4 | 7 | 13 | 25 | 7 | 19 | 37 | 73 | 145 | Integrable |
| 4 | | $00 \to 00, 01 \to 11, 10 \to 01, 11 \to 10$ | 2 | 4 | 7 | 13 | 25 | 7 | 19 | 37 | 73 | 145 | Integrable |
| 5 | East | $00 \to 00, 01 \to 11, 10 \to 10, 11 \to 01$ | 1 | 1 | 1 | 1 | 1 | 1 | 1 | 1 | 1 | 1 | Fully chaotic |
| 6 | Inverted West | $00 \to 01, 01 \to 00, 10 \to 10, 11 \to 11$ | 1 | 1 | 1 | 1 | 1 | 1 | 1 | 1 | 1 | 1 | Fully chaotic |
| 7 | | $00 \to 01, 01 \to 00, 10 \to 11, 11 \to 10$ | 1 | 5 | 17 | 65 | 257 | 13 | 49 | 193 | 769 | 3073 | Integrable |
| 8 | | $00 \to 01, 01 \to 10, 10 \to 00, 11 \to 11$ | 2 | 4 | 7 | 13 | 25 | 7 | 19 | 37 | 73 | 145 | Integrable |
| 9 | | $00 \to 01, 01 \to 10, 10 \to 11, 11 \to 00$ | 1 | 1 | 1 | 1 | 1 | 1 | 1 | 1 | 1 | 1 | Fully chaotic |
| 10 | | $00 \to 01, 01 \to 11, 10 \to 00, 11 \to 10$ | 3 | 5 | 9 | 17 | 33 | 13 | 25 | 49 | 97 | 193 | Integrable |
| 11 | | $00 \to 01, 01 \to 11, 10 \to 10, 11 \to 00$ | 2 | 4 | 7 | 13 | 25 | 7 | 19 | 37 | 73 | 145 | Integrable |
| 12 | | $00 \to 10, 01 \to 00, 10 \to 01, 11 \to 11$ | 2 | 4 | 7 | 13 | 25 | 7 | 19 | 37 | 73 | 145 | Integrable |
| 13 | | $00 \to 10, 01 \to 00, 10 \to 11, 11 \to 01$ | 3 | 5 | 9 | 17 | 33 | 13 | 25 | 49 | 97 | 193 | Integrable |
| 14 | Inverted East | $00 \to 10, 01 \to 01, 10 \to 00, 11 \to 11$ | 1 | 1 | 1 | 1 | 1 | 1 | 1 | 1 | 1 | 1 | Fully chaotic |
| 15 | | $00 \to 10, 01 \to 01, 10 \to 11, 11 \to 00$ | 2 | 4 | 7 | 13 | 25 | 7 | 19 | 37 | 73 | 145 | Integrable |
| 16 | | $00 \to 10, 01 \to 11, 10 \to 00, 11 \to 01$ | 1 | 5 | 17 | 65 | 257 | 13 | 49 | 193 | 769 | 3073 | Integrable |
| 17 | | $00 \to 10, 01 \to 11, 10 \to 01, 11 \to 00$ | 1 | 1 | 1 | 1 | 1 | 1 | 1 | 1 | 1 | 1 | Fully chaotic |
| 18 | | $00 \to 11, 01 \to 00, 10 \to 01, 11 \to 10$ | 1 | 1 | 1 | 1 | 1 | 1 | 1 | 1 | 1 | 1 | Fully chaotic |
| 19 | | $00 \to 11, 01 \to 00, 10 \to 10, 11 \to 01$ | 2 | 4 | 7 | 13 | 25 | 7 | 19 | 37 | 73 | 145 | Integrable |
| 20 | | $00 \to 11, 01 \to 01, 10 \to 00, 11 \to 10$ | 2 | 4 | 7 | 13 | 25 | 7 | 19 | 37 | 73 | 145 | Integrable |
| 21 | | $00 \to 11, 01 \to 01, 10 \to 10, 11 \to 00$ | 3 | 5 | 9 | 17 | 33 | 13 | 25 | 49 | 97 | 193 | Integrable |
| 22 | | $00 \to 11, 01 \to 10, 10 \to 00, 11 \to 01$ | 1 | 1 | 1 | 1 | 1 | 1 | 1 | 1 | 1 | 1 | Fully chaotic |
| 23 | NOT | $00 \to 11, 01 \to 10, 10 \to 01, 11 \to 00$ | 3 | 9 | 33 | 129 | 513 | 13 | 49 | 193 | 769 | 3073 | Integrable |

TABLE I. Exhaustive list of likely classification of all classical deterministic gates with local dimension $d = 2$. *Normal* CQs corresponds to conserved quantities with $\mu = 1$ and $\lambda = \pm 1$. *Staggered* CQs corresponds to ones where $\mu, \lambda$ are higher order roots of unity. We ran a search for $\lambda^n, \mu$ up to 12th roots of unity, $m, n \leq 12$. For the likely integrable gates, it may that more CQs exist at higher orders. Thus, the number only represents a lower bound. Note that there is always one CQ, corresponding to the flat state $\langle -| \langle -|$.

---

* swk34@cantab.ac.uk

## B. FINDING CONSERVED QUANTITIES AND GAUGE FIXING

The naive approach (Eq. (10)) will yield a myriad of solutions. This is because we have not considered gauge freedom. Note that due to the summation over space, for some solution $\langle f|$, $\langle f| + \mu^{-2}\langle \vartheta, -, -| - \langle -, -, \vartheta|$ for some gauge field $\langle \vartheta|$ of length $2(l-1)$, gives the same $\langle F|$. Here, we used a shorthand $\langle f, g| := \langle f| \langle g|$. Therefore any $\langle f|$ satisfying

$$\Big( \langle f_o, -| + \langle -, f_e| \Big) \mathbf{u}_{1:2l} = \Big( \mu^{-2} \langle f_e, -| + \langle -, f_o| \Big) \lambda \mu + \mu^{-2} \langle C, -, -| - \langle -, -, C|, \tag{B.1}$$

for any $\langle C|$ of length $2(l-1)$, results in a valid local CQ. There is one additional gauge freedom. For given any solution $\langle f_e|, \langle f_o|$, the gauge transformation

$$\langle f_o| \rightarrow \langle f_o'| = \langle f_o| + \langle -, \chi|, \qquad \langle f_e| \rightarrow \langle f_e'| = \langle f_e| - \langle \chi, -|, \tag{B.2}$$

for some gauge field $\langle \chi|$ of length $2(l-1)$, also gives the same $\langle F|$. The previous gauge freedom can be assigned to $\langle f_e|, \langle f_o|$, as

$$\langle f_o| \rightarrow \langle f_o'| = \langle f_o| + \mu^{-2}\langle \vartheta, -|, \qquad \langle f_e| \rightarrow \langle f_e'| = \langle f_e| - \langle -, \vartheta|. \tag{B.3}$$

We therefore have two gauge freedoms $\langle \vartheta|$ and $\langle \chi|$, each with $d^{2(l-1)}$ degrees of freedom, with a total of $d^{4(l-1)}$. Suppose that we have a solution obeying Eq. (B.1). We can first write $\langle f|$ in terms of $\langle f'|$ to cancel $\langle C|$. After this, any gauge that will not produce any $\langle C|$ term requires that $\langle \vartheta| \mathbf{u}_{1:2l-2} = -\lambda \mu \langle \chi|$. With the remaining $d^{2(l-1)}$ degrees of freedom, we set particular values of $\langle f|$ to zero for various cases. For example, for the case where $\mu \neq 1$, we set $\langle f|0,0,a_3,..,a_{2l}\rangle = 0 \ \forall (a_3,..,a_{2l}) \neq (0,..,0)$, and for the case where $\mu = 1$, we set $\langle f_e|0,\cdot\cdot,0\rangle = 0$.

For a given gate $\sigma$, we use the null space method, where we set $l$, $\mu$, and $\lambda$ and look for the null space of the condition Eq. (10) subject to additional gauge-fixing constraints. We start from $l = m = n = 1$ case then increase the search space for higher roots of unity to determine conserved quantities of higher periodicity and locality.

## C. CONSTRUCTING HYDRODYNAMIC EQUATIONS IN THE GENERAL CASE

Consider a species of conserved quantity labelled by $\alpha$. Eq. (10) can also be written as

$$\Big( \langle F_i^{o,(\alpha)}| + \langle F_{i+1}^{e,(\alpha)}| \Big) \mathbf{u}_1 = \lambda_\alpha \mu_\alpha \left( \langle F_i^{o,(\alpha)}| \mathbf{T}^{-1} + \frac{1}{\mu_\alpha^2} \langle F_{i+1}^{e,(\alpha)}| \mathbf{T} \right), \tag{C.1}$$

where $i$ is odd. Then for the time-evolved charge $\langle F_{i,\tau}^{(\alpha)}|$, where $\langle F_{i,0}^{(\alpha)}| = \langle F_i^{(\alpha)}| = \langle F_i^{o,(\alpha)}| + \langle F_{i+1}^{e,(\alpha)}|$, we have

$$\langle F_{i,2}^{(\alpha)}| - \lambda_\alpha^2 \langle F_{i,0}^{(\alpha)}| = \langle F_i^{(\alpha)}|\mathbb{U}_2 - \lambda_\alpha^2 \langle F_i^{(\alpha)}| \tag{C.2}$$

$$= \langle F_i^{(\alpha)}|\mathbf{u}_1 \mathbf{T}\mathbf{u}_1 \mathbf{T}^{-1} - \lambda_\alpha^2 \langle F_i^{(\alpha)}| \tag{C.3}$$

$$= \lambda_\alpha \mu_\alpha \left( \langle F_i^{o,(\alpha)}| \mathbf{T}^{-1} + \frac{1}{\mu_\alpha^2} \langle F_{i+1}^{e,(\alpha)}| \mathbf{T} \right) \mathbf{T}\mathbf{u}_1 \mathbf{T}^{-1} - \lambda_\alpha^2 \langle F_i^{(\alpha)}| \tag{C.4}$$

$$= \lambda_\alpha \mu_\alpha \left( \langle F_i^{o,(\alpha)}| + \frac{1}{\mu_\alpha^2} \left( \langle F_i^{(\alpha)}| - \langle F_i^{o,(\alpha)}| \right) \mathbf{T}^2 \right) \mathbf{u}_1 \mathbf{T}^{-1} - \lambda_\alpha^2 \langle F_i^{(\alpha)}|. \tag{C.5}$$

At this point we use the fact that $\mathbf{T}^2 \mathbf{u}_1 \mathbf{T}^{-2} = \mathbf{u}_1$ to simplify $\mathbf{T}^2 \mathbf{u}_1 \mathbf{T}^{-1} = \mathbf{u}_1 \mathbf{T}$, then use the identity Eq. (C.1) again on $\langle F_i^{(\alpha)}|$ to get

$$\langle F_{i,2}^{(\alpha)}| - \lambda_\alpha^2 \langle F_{i,0}^{(\alpha)}| = \lambda_\alpha \mu_\alpha \langle F_i^{o,(\alpha)}|\mathbf{u}_1 \mathbf{T}\mathbf{T}^{-2} + \frac{\lambda_\alpha^2}{\mu_\alpha^2} \langle F_{i+1}^{e,(\alpha)}|\mathbf{T}^2 - \frac{\lambda_\alpha}{\mu_\alpha} \langle F_i^{o,(\alpha)}|\mathbf{u}_1 \mathbf{T} - \lambda^2 \langle F_{i+1}^{e,(\alpha)}|. \tag{C.6}$$

Now defining "bare" current as

$$\langle G_i^{(\alpha)}| := -\lambda_\alpha \mu_\alpha^{-1} \langle F_i^{o,(\alpha)}| + \lambda^2 \mu^{-2} \langle F_{i+1}^{e,(\alpha)}|\mathbf{T}^{-2}, \tag{C.7}$$

we have

$$\langle F_{i,2}^{(\alpha)}| - \lambda_\alpha^2 \langle F_{i,0}^{(\alpha)}| = -\Big( \mu_\alpha^2 \langle G_{i+2,0}^{(\alpha)}| - \langle G_{i,0}^{(\alpha)}| \Big), \tag{C.8}$$

where $\langle G_{i+2}^{(\alpha)}| = \langle G_i^{(\alpha)}|\mathbf{T}^{-2}$.

Now we consider time evolution of charge defined in Eq. (16) by superunit of time $n$. Since by definition $\lambda_\alpha^{-n} = \lambda_\alpha^{-0} = 1$, we have

$$\langle q_{i,n}^{(\alpha)}| - \langle q_{i,0}^{(\alpha)}| = \lambda_\alpha^{-n}\langle q_{i,n}^{(\alpha)}| - \lambda_\alpha^{-0}\langle q_{i,0}^{(\alpha)}| \tag{C.9}$$

$$= \left(\lambda_\alpha^{-n}\langle q_{i,n}^{(\alpha)}| - \lambda_\alpha^{-(n-2)}\langle q_{i,n-2}^{(\alpha)}|\right) + \left(\lambda_\alpha^{-(n-2)}\langle q_{i,n-2}^{(\alpha)}| - \lambda_\alpha^{-(n-4)}\langle q_{i,n-4}^{(\alpha)}|\right) + \cdots + \left(\lambda_\alpha^{-2}\langle q_{i,2}^{(\alpha)}| - \lambda_\alpha^{-0}\langle q_{i,0}^{(\alpha)}|\right) \tag{C.10}$$

$$= \sum_{\tau=0,\text{ even}}^{n-2}\left(\lambda_\alpha^{-(\tau+2)}\langle q_{i,\tau+2}^{(\alpha)}| - \lambda_\alpha^{-\tau}\langle q_{i,\tau}^{(\alpha)}|\right). \tag{C.11}$$

Expanding $\langle q_{i,\tau}^{(\alpha)}|$,

$$\langle q_{i,n}^{(\alpha)}| - \langle q_{i,0}^{(\alpha)}| = \frac{1}{m}\sum_{\tau=0,\text{ even}}^{n-2}\sum_{j=0,\text{ even}}^{m-2}\left(\lambda_\alpha^{-(\tau+2)}\mu_\alpha^j\langle F_{i+j,\tau+2}^{(\alpha)}| - \lambda_\alpha^{-\tau}\mu_\alpha^j\langle F_{i+j,\tau}^{(\alpha)}|\right) \tag{C.12}$$

$$= \frac{1}{m}\sum_{\tau=0,\text{ even}}^{n-2}\sum_{j=0,\text{ even}}^{m-2}\lambda_\alpha^{-\tau}\mu_\alpha^j\left(\langle F_{i+j,\tau+2}^{(\alpha)}| - \lambda_\alpha^2\langle F_{i+j,\tau}^{(\alpha)}|\right) \tag{C.13}$$

$$= \frac{1}{m}\sum_{\tau=0,\text{ even}}^{n-2}\sum_{j=0,\text{ even}}^{m-2}\lambda_\alpha^{-\tau}\mu_\alpha^j\left(\langle F_{i+j,2}^{(\alpha)}| - \lambda_\alpha^2\langle F_{i+j,0}^{(\alpha)}|\right)\mathbf{U}^{\tau/2}. \tag{C.14}$$

Using the bare continuity relation Eq. (C.8), we have

$$\langle q_{i,n}^{(\alpha)}| - \langle q_{i,0}^{(\alpha)}| = -\frac{1}{m}\sum_{\tau=0,\text{ even}}^{n-2}\sum_{j=0,\text{ even}}^{m-2}\lambda_\alpha^{-\tau}\mu_\alpha^j\left(\mu_\alpha^2\langle G_{i+j+2,0}^{(\alpha)}| - \langle G_{i+j,0}^{(\alpha)}|\right)\mathbf{U}^{\tau/2} \tag{C.15}$$

$$= -\frac{1}{m}\sum_{\tau=0,\text{ even}}^{n-2}\sum_{j=0,\text{ even}}^{m-2}\lambda_\alpha^{-\tau}\mu_\alpha^j\left(\mu_\alpha^2\langle G_{i+j+2,0}^{(\alpha)}| - \langle G_{i+j,0}^{(\alpha)}|\right)\mathbf{U}^{\tau/2} \tag{C.16}$$

$$= -\frac{1}{m}\sum_{\tau=0,\text{ even}}^{n-2}\sum_{j=0,\text{ even}}^{m-2}\lambda_\alpha^{-\tau}\mu_\alpha^j\left(\mu_\alpha^2\langle G_{i+j+2,0}^{(\alpha)}| - \langle G_{i+j,0}^{(\alpha)}|\right)\mathbf{U}^{\tau/2} \tag{C.17}$$

$$= -\frac{1}{m}\left(\sum_{\tau=0,\text{ even}}^{n-2}\lambda_\alpha^{-\tau}\left(\mu_\alpha^m\langle G_{i+m,0}^{(\alpha)}| - \mu_\alpha^0\langle G_{i,0}^{(\alpha)}|\right)\mathbf{U}^{\tau/2}\right) \tag{C.18}$$

$$= -\frac{1}{m}\left(\sum_{\tau=0,\text{ even}}^{n-2}\lambda_\alpha^{-\tau}\left(\langle G_{i+m,0}^{(\alpha)}| - \langle G_{i,0}^{(\alpha)}|\right)\mathbf{U}^{\tau/2}\right). \tag{C.19}$$

Then, defining current as

$$\langle j_i^{(\alpha)}| = \frac{1}{n}\sum_{\tau=0,\text{ even}}^{n-2}\lambda_\alpha^{-\tau}\langle G_{i,0}^{(\alpha)}|\mathbf{U}^{\tau/2}, \tag{C.20}$$

we finally obtain the microscopy continuity Eq. (17) of the main text.

Next, we calculate the expectation on thermal states $|p_{\vec{\beta}}\rangle$. Note the two properties,

$$\mathbf{T}^2|p_{\vec{\beta}}\rangle = \frac{1}{Z(\vec{\beta})}\sum_{\boldsymbol{a}}\exp\left[-\sum_\alpha\beta_\alpha\langle F^{(\alpha)}|\boldsymbol{a}\rangle\right]\mathbf{T}^2|\boldsymbol{a}\rangle = \frac{1}{Z(\vec{\beta})}\sum_{\boldsymbol{a}}\exp\left[-\sum_\alpha\beta_\alpha\langle F^{(\alpha)}|\mathbf{T}^{-2}|\boldsymbol{a}\rangle\right]|\boldsymbol{a}\rangle \tag{C.21}$$

$$= \frac{1}{Z(\vec{\beta})}\sum_{\boldsymbol{a}}\exp\left[-\sum_\alpha\frac{\beta_\alpha}{\mu_\alpha^2}\langle F^{(\alpha)}|\boldsymbol{a}\rangle\right]|\boldsymbol{a}\rangle = |p_{\overrightarrow{(\beta/\mu)}}\rangle. \tag{C.22}$$

$$\mathbf{u}_1\mathbf{T}|p_{\vec{\beta}}\rangle = \frac{1}{Z(\vec{\beta})}\sum_{\boldsymbol{a}}\exp\left[-\sum_{\alpha}\beta_\alpha\langle F^{(\alpha)}|\mathbf{T}^{-1}\mathbf{u}_1^{-1}|\boldsymbol{a}\rangle\right]|\boldsymbol{a}\rangle = \frac{1}{Z(\vec{\beta})}\sum_{\boldsymbol{a}}\exp\left[-\sum_\alpha\frac{\beta_\alpha}{\mu_\alpha\lambda_\alpha}\langle F^{(\alpha)}|\boldsymbol{a}\rangle\right]|\boldsymbol{a}\rangle \tag{C.23}$$

$$= |p_{\overrightarrow{(\beta/\lambda\mu)}}\rangle. \tag{C.24}$$

Then the expectation value of the charge on the thermal state is

$$\langle q_i^{(\alpha)}\rangle_{\vec{\beta}} = \frac{1}{m}\sum_{j=0,\text{ even}}^{m-2}\mu_\alpha^j\langle F_{i+j}^{(\alpha)}\rangle_{\vec{\beta}} = \frac{1}{m}\sum_{j=0,\text{ even}}^{m-2}\mu_\alpha^j\langle F_i^{(\alpha)}\rangle_{\overrightarrow{(\beta\mu^j)}}. \tag{C.25}$$

The expectation of time-evolved bare current on thermal states is

$$\langle G_i^{(\alpha)}|\mathbf{U}^{\tau/2}|p_{\vec{\beta}}\rangle = \langle G_i^{(\alpha)}|(\mathbf{u}_1\mathbf{T}\mathbf{u}_1\mathbf{T}^{-1})^{\tau/2}|p_{\vec{\beta}}\rangle = \langle G_i^{(\alpha)}|p_{\overrightarrow{(\beta/\lambda^\tau)}}\rangle \tag{C.26}$$

$$= -\frac{\lambda_\alpha}{\mu_\alpha}\langle f_\text{o}^{(\alpha)}\rangle_{\overrightarrow{(\beta/\lambda^{\tau+1}\mu)}} + \frac{\lambda_\alpha^2}{\mu_\alpha^2}\langle f_\text{e}^{(\alpha)}\rangle_{\overrightarrow{(\beta/\lambda^\tau\mu^2)}} \tag{C.27}$$

Therefore the expectation of the current is

$$\langle \jmath_i^{(\alpha)}\rangle_{\vec{\beta}} = \frac{1}{n}\sum_{\tau=0,\text{ even}}^{n-2}\lambda_\alpha^{-\tau}\langle G_{i,0}^{(\alpha)}|\mathbf{U}^{\tau/2}|p_{\vec{\beta}}\rangle \tag{C.28}$$

$$= \frac{1}{n}\sum_{\tau=0,\text{ even}}^{n-2}\left(-\frac{1}{\lambda_\alpha^{\tau-1}\mu_\alpha}\langle f_\text{o}^{(\alpha)}\rangle_{\overrightarrow{(\beta/\lambda^{\tau+1}\mu)}} + \frac{1}{\lambda_\alpha^{\tau-2}\mu_\alpha^2}\langle f_\text{e}^{(\alpha)}\rangle_{\overrightarrow{(\beta/\lambda^\tau\mu^2)}}\right). \tag{C.29}$$

Finally Eq. (19) of main text can be resolved as

$$\partial_t\vec{q} = -\vec{B}\vec{C}^{-1}\partial_x\vec{q}, \tag{C.30}$$

where $[\vec{B}]_{\alpha,\alpha'} = \partial\jmath^{(\alpha)}/\partial\beta_{\alpha'}$ and $[\vec{C}]_{\alpha,\alpha'} = \partial q^{(\alpha)}/\partial\beta_{\alpha'}$.

## D. DECOUPLING OF STAGGERED CQS FROM SIMPLE CQS

Now, we would like to consider a special case where conserved quantities are either non-staggered both in space and time, i.e. $\lambda_\alpha^2 = \mu_\alpha^2 = 1$, or staggered in space and time $\lambda_\alpha^2 \neq 1, \mu_\alpha^2 \neq 1$. For sake of slim notation, we will label by $\alpha$ a conserved quantity of the first type and by $\bar{\alpha}$ a conserved quantity of the second type. We want to show that in this case, whenever $\beta_{\bar{\alpha}} = 0$ for all $\bar{\alpha}$ the hydrodynamic equations for the non-staggered conserved quantities decouple from the staggered ones.

First, note that in this case we have

$$\overrightarrow{(\beta\lambda^\tau\mu^j)} = \vec{\beta}, \tag{D.1}$$

which inserted into (C.25) immediately gives

$$\langle q_i^{(\bar{\alpha})}\rangle_{\vec{\beta}} = \frac{1}{m}\sum_{j=0,\text{ even}}^{m-2}\mu_{\bar{\alpha}}^j\langle F_i^{(\bar{\alpha})}\rangle_{\vec{\beta}} = 0. \tag{D.2}$$

This cancellation happens since $\mu_{\bar{\alpha}}$ is an $m_{\bar{\alpha}}$ root of unity. Therefore, $\beta_{\bar{\alpha}} = 0$ corresponds to $\langle q_i^{(\bar{\alpha})}\rangle_{\vec{\beta}} = 0$.

Similarly, we can compute the two matrices $\vec{C}$ and $\vec{B}$ and find that they are block-diagonal in $\alpha, \bar{\alpha}$:

$$[\vec{C}]_{\alpha,\bar{\alpha}} = \frac{\mathrm{d}}{\mathrm{d}\beta_{\bar{\alpha}}}\langle q_i^{(\alpha)}\rangle_{\vec{\beta}}\bigg|_{\beta_{\bar{\alpha}}=0} = \frac{1}{m}\sum_{j=0,\text{ even}}^{m-2}\mu_{\bar{\alpha}}^j\frac{\mathrm{d}}{\mathrm{d}\beta_{\bar{\alpha}}}\langle F_i^{(\alpha)}\rangle_{\vec{\beta}}\bigg|_{\beta_{\bar{\alpha}}=0} = 0 \tag{D.3}$$

$$[\vec{B}]_{\alpha,\bar{\alpha}} = \frac{\mathrm{d}}{\mathrm{d}\beta_{\bar{\alpha}}}\langle\jmath_i^{(\alpha)}\rangle_{\vec{\beta}}\bigg|_{\beta_{\bar{\alpha}}=0} = \frac{1}{n}\sum_{\tau=0,\text{ even}}^{n-2}\left(-\frac{1}{\lambda_{\bar{\alpha}}^{\tau-1}\mu_{\bar{\alpha}}}\frac{\mathrm{d}}{\mathrm{d}\beta_{\bar{\alpha}}}\langle f_\text{o}^{(\alpha)}\rangle_{\vec{\beta}}\bigg|_{\beta_{\bar{\alpha}}=0} + \frac{1}{\lambda_{\bar{\alpha}}^{\tau-2}\mu_{\bar{\alpha}}^2}\frac{\mathrm{d}}{\mathrm{d}\beta_{\bar{\alpha}}}\langle f_\text{e}^{(\alpha)}\rangle_{\vec{\beta}}\bigg|_{\beta_{\bar{\alpha}}=0}\right) = 0, \tag{D.4}$$

Since these two matrices are block-diagonal, also $\vec{A} = \vec{B}\vec{C}^{-1}$ is block-diagonal. Therefore the hydrodynamic equations read:

$$\partial_t q^{(\alpha)} + \sum_{\alpha'}[\vec{A}]_{\alpha,\alpha'}\partial_x q^{(\alpha')} = 0 \qquad\qquad \partial_t q^{(\bar{\alpha})} + \sum_{\bar{\alpha}'}[\vec{A}]_{\bar{\alpha},\bar{\alpha}'}\partial_x q^{(\bar{\alpha}')} = 0. \tag{D.5}$$

Recall that initially $q^{(\bar{\alpha})}(t = 0, x) = 0$ and therefore the second equation is trivially solved by $q^{(\bar{\alpha})}(t, x) = 0$ for all times. Thus, our assumption of $\beta_{\bar{\alpha}} = 0$ remains valid throughout the hydrodynamic evolution and therefore the hydrodynamic equations decouple at all times.

We would like to stress that this decoupling only applies when $\beta_{\bar{\alpha}} = 0$, but not for generic values. In fact, computing higher order derivatives of $\langle q_i^{(\alpha)} \rangle_{\vec{\beta}}$ and $\langle \jmath_i^{(\alpha)} \rangle_{\vec{\beta}}$ at $\beta_{\bar{\alpha}} = 0$ one can see that they will eventually couple between $\alpha$ and $\bar{\alpha}$. For instance if $\lambda_{\bar{\alpha}} = \mu_{\bar{\alpha}} = i$, both $\frac{\mathrm{d}^2}{\mathrm{d}\beta_{\bar{\alpha}}^2}\langle q_i^{(\alpha)} \rangle_{\vec{\beta}}\big|_{\beta_{\bar{\alpha}}=0} \neq 0$ and $\frac{\mathrm{d}^2}{\mathrm{d}\beta_{\bar{\alpha}}^2}\langle \jmath_i^{(\alpha)} \rangle_{\vec{\beta}}\big|_{\beta_{\bar{\alpha}}=0} \neq 0$, signaling a coupling between both sets of conserved quantities.

### E. DETAILS ON KARDAR-PARISI-ZHANG SCALING OF CORRELATIONS

Here we give more details on how we compare the correlation functions to the non-linear fluctuating hydrodynamic theory of [1] to predict not only KPZ superdiffusive scaling, but also the superdiffusion constant $\lambda_{\mathrm{B}}$.

Our starting point is Eq. (29). Defining $\phi(x, t) = \delta q(x, t)/\sqrt{C}$, where

$$C = \int \mathrm{d}x \, \langle \delta q(x, 0)\delta q(0, 0) \rangle_{\beta}^{\mathrm{c}} \tag{E.1}$$

and neglecting higher order terms we find

$$\partial_t \phi = -\partial_x \left( v(q_0)\phi + \frac{1}{2}\jmath''(q_0)\sqrt{C}\phi^2 \right) \tag{E.2}$$

with initial fluctuations satisfying $\langle \phi(0, x) \rangle_{\beta} = 0$ and $\langle \phi(x, 0)\phi(y, 0) \rangle_{\beta}^{\mathrm{c}} = \delta(x - y)$. By mapping this equation onto the KPZ equation and adding diffusion and noise terms, in [1] it was shown that the time-evolved correlations are given by

$$\langle \phi(x, t)\phi(0, 0) \rangle_{\beta}^{\mathrm{c}} = \frac{1}{(\lambda_{\mathrm{B}}t)^{2/3}} f_{\mathrm{KPZ}}\left( \frac{x - v(q_0)t}{(\lambda_{\mathrm{B}}t)^{2/3}} \right), \tag{E.3}$$

where $\lambda_{\mathrm{B}} = \sqrt{2C}|\jmath''(q_0)|$ and $f_{\mathrm{KPZ}}(x)$ is specified in [1, 2]. From this we find

$$\langle q(x, t)q(0, 0) \rangle_{\beta}^{\mathrm{c}} = \frac{C}{(\lambda_{\mathrm{B}}t)^{2/3}} f_{\mathrm{KPZ}}\left( \frac{x - v(q_0)t}{(\lambda_{\mathrm{B}}t)^{2/3}} \right). \tag{E.4}$$

In order to compare this theory with numerical simulations we need to compute the correct normalization $C$ from the microscopic data. This can be done as follows

$$C = \int \mathrm{d}x \, \langle q(x, 0)q(0, 0) \rangle_{\beta}^{\mathrm{c}} = \frac{1}{L\Delta x}\langle Q^2 \rangle_{\beta}^{\mathrm{c}} = \frac{1}{L\Delta x}\sum_{i,j \text{ odd}} \langle q_{i:i+1}q_{j:j+1} \rangle_{\beta}^{\mathrm{c}} \tag{E.5}$$

$$= \frac{1}{L\Delta x}\sum_{i \text{ odd}} \langle q_{i:i+1}^2 \rangle_{\beta}^{\mathrm{c}} = \frac{1}{2\Delta x}\langle q_{1:2}^2 \rangle_{\beta}^{\mathrm{c}}. \tag{E.6}$$

Here we used $Q = \int \mathrm{d}x \, q(x, 0) = \sum_{i \text{ odd}} q_{i:i+1}$ and the translation invariance of the state. Using this we can rewrite Eq. (30) into a prediction after applying $\tau$ layers:

$$\langle q_{i:i+1}(\tau\Delta t)q_{1:2}(0) \rangle_{\beta}^{\mathrm{c}} = \frac{2\Delta x \langle q_{1:2}(0)^2 \rangle_{\beta}^{\mathrm{c}}}{(\lambda_{\mathrm{B}}\tau\Delta t)^{2/3}} f_{\mathrm{KPZ}}\left( \frac{\Delta x i - v(q_0)\tau\Delta t}{(\lambda_{\mathrm{B}}\tau\Delta t)^{2/3}} \right). \tag{E.7}$$

Using the same units for space and time $\Delta x = \Delta t = 1$, as used throughout the paper, we finally find

$$\langle q_{i:i+1}(\tau)q_{j:j+1}(0) \rangle_{\beta}^{\mathrm{c}} = \frac{2 \langle q_{1:2}(0)^2 \rangle_{\beta}^{\mathrm{c}}}{(\lambda_{\mathrm{B}}\tau)^{2/3}} f_{\mathrm{KPZ}}\left( \frac{i - v(q_0)\tau}{(\lambda_{\mathrm{B}}\tau)^{2/3}} \right). \tag{E.8}$$

In particular we find that

$$\max_i \langle q_{i:i+1}(\tau\Delta t) q_{j:j+1}(0)\rangle_\beta^c = \frac{2\langle q_{1:2}(0)^2\rangle_\beta^c}{(\lambda_B \tau)^{2/3}} f_{KPZ}(0),$$ (E.9)

with $f_{KPZ}(0) \approx 0.54$. By fitting this function to the numerically observed maximum of the correlation function as function of $\tau$, we can extract the superdiffusion constant $\lambda_B$.

## F. CONSIDERATIONS ON CIRCUIT $(d, \sigma) = (3, 2312)$

In this section we give further details on the dynamics of $(d, \sigma) = (3, 2312)$, in particular the evolution of an individual $|2\rangle$. This is a good model for $\beta \to \infty$, where only very few $|2\rangle$s are emerged into a background of independently chosen $|0\rangle$s and $|1\rangle$s.

Note that the model is parity symmetric. Hence we can assume the $|2\rangle$ to be initially on an odd site, say $a_1 = 2$. Now consider the case where $a_0 = 1$: If $a_{-1} = 0$ we have the following evolution

$$|\ldots a_{-1}a_0 a_1 \ldots\rangle = |\ldots 012\ldots\rangle \to |\ldots 020\ldots\rangle \to |\ldots 120\ldots\rangle \to |\ldots 121\ldots\rangle$$ (F.1)

and thus after 3 steps we find again the combination $|\ldots 12\ldots\rangle$, but shifted one site to the left. Similarly if $a_{-1} = 1$ we have

$$|\ldots a_{-1}a_0 a_1 \ldots\rangle = |\ldots 112\ldots\rangle \to |\ldots 120\ldots\rangle.$$ (F.2)

and thus we find $|\ldots 12\ldots\rangle$ already after 1 step, shifted to the left. Since the probability of $|0\rangle$s and $|1\rangle$s is equal, we therefore find that on average the 2 moves one step to the left every 2 steps, i.e. its velocity is $-1/2$.

Now let us consider the missing case when $a_1 = 2$ and $a_0 = 0$. If $a_2 = 0$ we find

$$|\ldots a_0 a_1 a_2 \ldots\rangle = |\ldots 020\ldots\rangle = |\ldots 120\ldots\rangle \to |\ldots 121\ldots\rangle,$$ (F.3)

at which point we identify again the 12 combination, implying that the 2 will be a left mover. In the other case $a_2 = 1$ we have

$$|\ldots a_0 a_1 a_2 \ldots\rangle = |\ldots 021\ldots\rangle = |\ldots 121\ldots\rangle,$$ (F.4)

from which we can read off that the next gate will apply to $|21\rangle$, hence by parity symmetry this $|2\rangle$ will be a right mover. We conclude that single $|2\rangle$'s are either left or right movers with an average velocity $1/2$.

If there are multiple $|2\rangle$s we can say that the total number of left and right movers

$$N_L = \sum_{i \text{ even}} \delta_{a_i 1}\delta_{a_{i+1} 2} + \delta_{a_i 0}\delta_{a_{i+1} 2}\delta_{a_{i+2} 0} + \delta_{a_{i-1} 1}\delta_{a_i 2}\delta_{a_{i+1} 0}$$ (F.5)

$$N_R = \sum_{i \text{ even}} \delta_{a_i 2}\delta_{a_{i+1} 1} + \delta_{a_i 0}\delta_{a_{i+1} 2}\delta_{a_{i+2} 1} + \delta_{a_{i-1} 0}\delta_{a_i 2}\delta_{a_{i+1} 0}$$ (F.6)

are approximately conserved. Under two-body scattering the number of left and right movers seems to be preserved (although we cannot prove it), so that these quantities are only non-conserved by the less frequent three-body (or $n$-body) scattering events. Therefore, we expect that $N_L$ and $N_R$ can be treated as approximately conserved for short times, in particular in the low density regime.

---



\end{document}

%% file: circuits.tex
\usepackage{tikz} 
\usepackage{xstring} 
\usepackage{relsize} 

\usetikzlibrary{decorations.markings}
\usetikzlibrary{snakes}
\usetikzlibrary{shapes.geometric}
\tikzset{fontscale/.style = {font=\relsize{#1}}}

\NewDocumentCommand{\squircle}{ O{.0} O{.0} }{
    \def\qs{.25}
    \draw[thick, fill=white, rounded corners=1]
    ({(#1)-\qs},{(#2)-\qs}) -- ({(#1)-1/2},{(#2)-1/2})
    ({(#1)+\qs},{(#2)-\qs}) -- ({(#1)+1/2},{(#2)-1/2})
    ({(#1)-\qs},{(#2)+\qs}) -- ({(#1)-1/2},{(#2)+1/2})
    ({(#1)+\qs},{(#2)+\qs}) -- ({(#1)+1/2},{(#2)+1/2})
    ({(#1)-\qs},{(#2)-\qs}) rectangle ({(#1)+\qs},{(#2)+\qs});
}

\NewDocumentCommand{\giyeok}{  O{TR} O{.0} O{.0} }{
    \def\gs{.19}
    \IfEqCase{#1}{
        {TR}{\draw[thick, rounded corners=1]({(#2)+.0},{(#3)+\gs}) -- ({(#2)+\gs},{(#3)+\gs}) -- ({(#2)+\gs},{(#3)+.0});}
        {TL}{\draw[thick, rounded corners=1]({(#2)+.0},{(#3)+\gs}) -- ({(#2)-\gs},{(#3)+\gs}) -- ({(#2)-\gs},{(#3)+.0});}
        {BR}{\draw[thick, rounded corners=1]({(#2)+\gs},{(#3)+.0}) -- ({(#2)+\gs},{(#3)-\gs}) -- ({(#2)+.0},{(#3)-\gs});}
        {BL}{\draw[thick, rounded corners=1]({(#2)-\gs},{(#3)+.0}) -- ({(#2)-\gs},{(#3)-\gs}) -- ({(#2)+.0},{(#3)-\gs});}
    }
}

\NewDocumentCommand{\stub}{ O{D} O{.0} O{.0} m}{
    \IfEqCase{#1}{
        {D}{
            \IfEqCase{#4}{
                {TR}{\draw[thick](#2+.5-.15,#3+.5+.15) -- (#2+.5+.15,#3+.5-.15);}
                {TL}{\draw[thick](#2-.5-.15,#3+.5-.15) -- (#2-.5+.15,#3+.5+.15);}
                {BR}{\draw[thick](#2+.5-.15,#3-.5-.15) -- (#2+.5+.15,#3-.5+.15);}
                {BL}{\draw[thick](#2-.5-.15,#3-.5+.15) -- (#2-.5+.15,#3-.5-.15);}
            }
        }
        {H}{
            \IfEqCase{#4}{
                {TR}{\draw[thick](#2+.5-.15,#3+.5) -- (#2+.5+.15,#3+.5);}
                {TL}{\draw[thick](#2-.5-.15,#3+.5) -- (#2-.5+.15,#3+.5);}
                {BR}{\draw[thick](#2+.5-.15,#3-.5) -- (#2+.5+.15,#3-.5);}
                {BL}{\draw[thick](#2-.5-.15,#3-.5) -- (#2-.5+.15,#3-.5);}
            }
        }
        {V}{
            \IfEqCase{#4}{
                {TR}{\draw[thick](#2+.5,#3+.5+.15) -- (#2+.5,#3+.5-.15);}
                {TL}{\draw[thick](#2-.5,#3+.5-.15) -- (#2-.5,#3+.5+.15);}
                {BR}{\draw[thick](#2+.5,#3-.5-.15) -- (#2+.5,#3-.5+.15);}
                {BL}{\draw[thick](#2-.5,#3-.5+.15) -- (#2-.5,#3-.5-.15);}
            }
        }
    }
}

\NewDocumentCommand{\state}{ O{} O{.0} O{.0} O{0.35} m}{
\IfEqCase{#5}{
    {TR}{\draw[thick, fill=white]({(#2)+.5},{(#3)+.5+.5*(#4)}) circle [radius=(#4)] node {\small #1};}
    {TL}{\draw[thick, fill=white]({(#2)-.5},{(#3)+.5+.5*(#4)}) circle [radius=(#4)] node {\small #1};}
    {BR}{\draw[thick, fill=white]({(#2)+.5},{(#3)-.5-.5*(#4)}) circle [radius=(#4)] node {\small #1};}
    {BL}{\draw[thick, fill=white]({(#2)-.5},{(#3)-.5-.5*(#4)}) circle [radius=(#4)] node {\small #1};}
    {TRD}{\draw[thick, fill=white]({(#2)+.5+.5*(#4)},{(#3)+.5+.5*(#4)}) circle [radius=(#4)] node {\small #1};}
    {TLD}{\draw[thick, fill=white]({(#2)-.5-.5*(#4)},{(#3)+.5+.5*(#4)}) circle [radius=(#4)] node {\small #1};}
    {BRD}{\draw[thick, fill=white]({(#2)+.5+.5*(#4)},{(#3)-.5-.5*(#4)}) circle [radius=(#4)] node {\small #1};}
    {BLD}{\draw[thick, fill=white]({(#2)-.5-.5*(#4)},{(#3)-.5-.5*(#4)}) circle [radius=(#4)] node {\small #1};}
}
}

\NewDocumentCommand{\sqate}{ O{} O{.0} O{.0} O{0.35} m}{
\IfEqCase{#5}{
    {TR}{\draw[thick, fill=white]({(#2)+.5},{(#3)+.5+.5*(#4)}) circle [radius=(#4)] node {\small #1};}
    {TL}{\draw[thick, fill=white]({(#2)-.5},{(#3)+.5+.5*(#4)}) circle [radius=(#4)] node {\small #1};}
    {BR}{\draw[thick, fill=white]({(#2)+.5},{(#3)-.5-.5*(#4)}) circle [radius=(#4)] node {\small #1};}
    {BL}{\draw[thick, fill=white]({(#2)-.5},{(#3)-.5-.5*(#4)}) circle [radius=(#4)] node {\small #1};}
    {TRD}{\draw[thick, fill=white]({(#2)+.5-.1},{(#3)+.5-.1}) rectangle ++(#4*2, #4*2);}
    {TLD}{\draw[thick, fill=white]({(#2)-.5+.1},{(#3)+.5-.1}) rectangle ++(-#4*2, #4*2);}
    {BRD}{\draw[thick, fill=white]({(#2)+.5-.1},{(#3)-.5+.1}) rectangle ++(#4*2, -#4*2);}
    {BLD}{\draw[thick, fill=white]({(#2)-.5+.1},{(#3)-.5+.1}) rectangle ++(-#4*2, -#4*2);}
}
}

\NewDocumentCommand{\twostate}{ O{} O{.0} O{.0} m}{
    \begin{pgfinterruptboundingbox}
        \IfEqCase{#4}{
            {T}{\draw[thick, rounded corners=1, fill=white]({(#2)-.5-.1},{(#3)+.4}) rectangle node {\small #1} ++ (1.2, .65);}
            {B}{\draw[thick, rounded corners=1, fill=white]({(#2)-.5-.1},{(#3)-.4}) rectangle node {\small #1} ++ (1.2, .65);}
            {L}{\draw[thick, rounded corners=1, fill=white]({(#2)-.5-.5},{(#3)-.6}) rectangle node {\small #1} ++ (.65, 1.2);}
            {R}{\draw[thick, rounded corners=1, fill=white]({(#2)+.5},{(#3)-.6}) rectangle node {\small #1} ++ (.65, 1.2);}
        }
    \end{pgfinterruptboundingbox}
}

\NewDocumentCommand{\proj}{ O{} O{.0} O{.0} m}{
\IfEqCase{#4}{
    {T}{\draw[thick, fill=white]({(#2)},{(#3)+.5}) +(-0.175,-0.175) rectangle node {\tiny #1} +(0.175,0.175);}
    {B}{\draw[thick, fill=white]({(#2)},{(#3)+.5}) +(-0.175,-0.175) rectangle node {\tiny #1} +(0.175,0.175);}
    {L}{\draw[thick, fill=white]({(#2)-.5},{(#3)}) +(-0.175,-0.175) rectangle node {\tiny #1} +(0.175,0.175);}
    {R}{\draw[thick, fill=white]({(#2)+.5},{(#3)}) +(-0.175,-0.175) rectangle node {\tiny #1} +(0.175,0.175);}
}
}

\NewDocumentCommand{\tria}{ O{} O{.0} O{.0} m}{
\IfEqCase{#4}{
    {T}{\draw[thick, fill=white]({(#2)},{(#3)+.5}) +(-0.175,-0.175) -- +(+0.175,-0.175) -- +(0.0,0.175) -- cycle node {\tiny #1};}
    {B}{\draw[thick, fill=white]({(#2)},{(#3)+.5}) +(-0.175,-0.175) -- +(+0.175,-0.175) -- +(0.0,0.175) -- cycle node {\tiny #1};}
    {L}{\draw[thick, fill=white]({(#2)-.5},{(#3)}) +(-0.175,-0.175) -- +(+0.175,-0.175) -- +(0.0,0.175) -- cycle node {\tiny #1};}
    {R}{\draw[thick, fill=white]({(#2)+.5},{(#3)}) +(-0.175,-0.175) -- +(+0.175,-0.175) -- +(0.0,0.175) -- cycle node {\tiny #1};}
}
} 

\newcommand\inlinetikz[1]{
    \begin{tikzpicture}[baseline={([yshift=-.5ex] current bounding box.center)},scale=0.75]
        {#1}
    \end{tikzpicture}
}

\NewDocumentCommand{\gates}{ O{1} O{1} m }{
    \inlinetikz{
        \draw[thick, opacity=.0](-.9, -.9) rectangle (#1-1+.9, #2-1+.9);
        \grid{#1}{#2}
        {#3}
    }
}

\newcommand\gate[1]{
    \gates[1][1]{#1}
}

\NewDocumentCommand{\leg}{ O{.0} O{.0} m }{
    \IfEqCase{#3}{
        {T}{\draw[thick, rounded corners=1](#1-.5,#2+.5) -- (#1+.5,#2+.5);}
        {B}{\draw[thick, rounded corners=1](#1-.5,#2-.5) -- (#1+.5,#2-.5);}
        {R}{\draw[thick, rounded corners=1](#1+.5,#2-.5) -- (#1+.5,#2+.5);}
        {L}{\draw[thick, rounded corners=1](#1-.5,#2-.5) -- (#1-.5,#2+.5);}
        {D}{\draw[thick, rounded corners=1](#1-.5,#2-.5) -- (#1+.5,#2+.5);}
        {-D}{\draw[thick, rounded corners=1](#1+.5,#2-.5) -- (#1-.5,#2+.5);}
        {-DM}{\draw[thick, rounded corners=1](#1+.5,#2-.5) -- (#1-.5,#2);}
    }
}

\NewDocumentCommand{\arrowleg}{ O{.0} O{.0} m }{
    \IfEqCase{#3}{
        {T}{\draw[->-, thick, rounded corners=1](#1-.5,#2+.5) -- (#1+.5,#2+.5);}
        {B}{\draw[->-, thick, rounded corners=1](#1-.5,#2-.5) -- (#1+.5,#2-.5);}
        {R}{\draw[->-, thick, rounded corners=1](#1+.5,#2-.5) -- (#1+.5,#2+.5);}
        {L}{\draw[->-, thick, rounded corners=1](#1-.5,#2-.5) -- (#1-.5,#2+.5);}
        {D}{\draw[->-, thick, rounded corners=1](#1-.5,#2-.5) -- (#1+.5,#2+.5);}
        {-D}{\draw[->-, thick, rounded corners=1](#1+.5,#2-.5) -- (#1-.5,#2+.5);}
        {-DM}{\draw[->-, thick, rounded corners=1](#1+.5,#2-.5) -- (#1-.5,#2);}
    }
}

\NewDocumentCommand{\squiggly}{ O{.0} O{.0} m }{
    \tikzset{decoration={snake,amplitude=0.5, segment length=2}} 
    \IfEqCase{#3}{
        {T}{\draw[decorate, thick](#1+.5,#2+.5) arc[start angle=0, end angle=-180, radius=.5];}
        {B}{\draw[decorate, thick](#1+.5,#2-.5) arc[start angle=0, end angle=180, radius=.5];}
        {L}{\draw[decorate, thick](#1-.5,#2-.5) arc[start angle=-90, end angle=90, radius=.5];}
        {R}{\draw[decorate, thick](#1+.5,#2+.5) arc[start angle=90, end angle=270, radius=.5];}
    }
}

\NewDocumentCommand{\arcy}{ O{.0} O{.0} m }{
    \IfEqCase{#3}{
        {T}{\draw[thick](#1+.5,#2+.5) arc[start angle=0, end angle=-180, radius=.5];}
        {B}{\draw[thick](#1+.5,#2-.5) arc[start angle=0, end angle=180, radius=.5];}
        {L}{\draw[thick](#1-.5,#2-.5) arc[start angle=-90, end angle=90, radius=.5];}
        {R}{\draw[thick](#1+.5,#2+.5) arc[start angle=90, end angle=270, radius=.5];}
    }
}

\NewDocumentCommand{\writeat}{ O{.0} O{.0} m m}{
    \begin{pgfinterruptboundingbox}
    \IfEqCase{#3}{
        {M}{\node[] at ({(#1)}, {(#2)}) {#4};}
        {B}{\node[] at ({(#1)}, {(#2)-1.}) {#4};}
        {BL}{\node[] at ({(#1)-.8}, {(#2)-.8}) {#4};}
        {BR}{\node[] at ({(#1)+.8}, {(#2)-.8}) {#4};}
        {BLT}{\node[] at ({(#1)-.5}, {(#2)-1.0}) {#4};}
        {BRT}{\node[] at ({(#1)+.5}, {(#2)-1.0}) {#4};}
        {TL}{\node[] at ({(#1)-.8}, {(#2)+.8}) {#4};}
        {TR}{\node[] at ({(#1)+.8}, {(#2)+.8}) {#4};}
        {TLT}{\node[] at ({(#1)-.5}, {(#2)+1.0}) {#4};}
        {TRT}{\node[] at ({(#1)+.5}, {(#2)+1.0}) {#4};}
    }
    \end{pgfinterruptboundingbox}
}

\newcommand\grid[2]{
    \def\X{#1}
    \def\Y{#2}
    \draw[thick, dotted, lightgray]
    \foreach \y in {0,...,\Y}{
      (-.5,\y-.5) -- (\X-1+.5,\y-.5)
    }
    \foreach \x in {0,...,\X}{
      (\x-.5,-.5) -- (\x-.5,\Y-.5)
    };
}

\NewDocumentCommand{\halfgrid}{ O{0} O{0} m m}{
    \def\X{#3}
    \def\Y{#4}
    \draw[thick, dotted, lightgray]
    \foreach \y in {#2,...,\Y}{
      (2*#1-.5,2*\y-.5) -- (2*\X-.5,2*\y-.5)
    }
    \foreach \x in {#1,...,\X}{
      (2*\x-.5,2*#2-.5) -- (2*\x-.5,2*\Y-.5)
    };
}

\NewDocumentCommand{\halfwriteat}{ O{0.} O{0.} m}{
    \node[] at ({2.*(#1)+.5}, {2.*(#2)+.5}) {#3};
}

\NewDocumentCommand{\halfgates}{ O{0.} O{0.} m}{
    \squircle[2*(#1)][2*(#2)] \giyeok[TR][2*(#1)][2*(#2)] \writeat[2*(#1)][2*(#2)]{M}{#3}
    \squircle[2*(#1)+1][2*(#2)+1] \giyeok[TR][2*(#1)+1][2*(#2)+1] \writeat[2*(#1)+1][2*(#2)+1]{M}{#3}
}

\tikzset{->-/.style={decoration={
  markings,
  mark=at position .5 with {\arrow{>}}},postaction={decorate}}}

